\documentclass[]{aa}  
\usepackage{hyperref}
\hypersetup{
    colorlinks=true,
    citecolor=blue,
    linkcolor=blue,
    urlcolor=blue,
}

\usepackage{graphicx}
\usepackage{txfonts}
\begin{document} 

\title{Equilibrium tides and magnetic activity \\ in stars with close-by massive planets}
%\title{Equilibrium tides and overshooting convection in stars with close-by massive planets}

%   \title{Equilibrium tides raised by massive hot Jupiters \\ and interface  dynamos in their late-type host stars}

%   \subtitle{An application to the intriguing case of WASP-18}
\subtitle{The intriguing case of WASP-18}

   \author{A.~F.~Lanza
%          \inst{1}
          \and
          S.~N.~Breton %\inst{1}
          }

   \institute{INAF-Osservatorio Astrofisico di Catania, Via S.~Sofia, 78 - I-95123 Catania, Italy \\
              \email{antonino.lanza@inaf.it, sylvain.breton@inaf.it}
 %        \and
 %            Universit\'e Paris-Cit\'e, Universit\'e Paris-Saclay, CEA, CNRS, AIM, F-91191 Gif-Sur-Yvette, France \\
 %            \email{sylvain.breton@inaf.it}
             }

   \date{Received ... ; accepted ...}
\titlerunning{Equilibrium tides and stellar magnetic activity}
% \abstract{}{}{}{}{} 
% 5 {} token are mandatory
 
  \abstract
   {}
   {WASP-18 is an F6V star that hosts a planet with a mass of $\sim 10$ Jupiter masses and an orbital period of $\sim 0.94$~days. In spite of its relatively fast rotation and young age, the star remains undetected in X-rays, thus implying a very low level of magnetic activity. To account for such unexpected properties,  we propose a mechanism that  modifies the internal stratification and the photospheric magnetic activity of a late-type main sequence star with a close-by massive planet based on the action of the equilibrium tide. }
   {{{ We speculate that}} the horizontal flow produced by the equilibrium tide may interact with the convective plumes in the overshoot layer below the stellar outer convective envelope. The interaction is characterised by a very high Reynolds number ($Re \sim 10^{10}$), leading to the development of turbulent boundary layers at the surface of such structures, whereas turbulent wakes extend over most of the overshoot layer that they straddle.  }
   {We propose that such a tidally induced  turbulence can lead to a reduction of the filling factor of the downdrafts in the overshoot layer. As a consequence, the absolute value of the {{sub-adiabatic}} gradient increases in that layer hindering the emergence of magnetic flux tubes responsible for the formation of photospheric starspots. We{{ conjecture}} that this process is occurring in WASP-18, thus providing a possible mechanism to account for the very low level of magnetic activity observed
 for such a planet host. }
   {}

   \keywords{planetary systems -- planet-star interactions -- Stars: solar-type -- Stars: activity -- Stars: magnetic fields -- Stars: individual: WASP-18 (HD~10069), WASP-12}

   \maketitle
%
%-------------------------------------------------------------------

\section{Introduction}
\label{introduction}
Late-type main sequence (MS) stars have an outer convective envelope surrounding a radiative zone. The convective motions in the envelope do not stop at the level where the Schwarzschild criterion for the onset of convection is marginally satisfied, but they overshoot in the radiative zone below because of the inertia of the downwardly directed fluid parcels \citep[e.g.][]{Zahn91,Kippenhahnetal13}. Convective motions in stellar outer convection zones develop over several pressure scale heights, $H_{\rm p}$, and are characterised by extremely large Rayleigh and Reynolds numbers of the order of at least of $10^{25}$ and $10^{12}-10^{13}$, respectively \citep{Priest84}. In such a regime, a strong asymmetry arises between upward and downward convective motions, with the latter taking the form of concentrated downdrafts extending over most of the depth of the convection zone and reaching the lower boundary with a velocity well in excess of that predicted by  the mixing-length theory usually adopted to model stellar convection. These downdrafts occupy an area of  the order of 10\% at the base of the convective envelope, while the upward directed motions are characterised by a wider areal filling factor and lower upward velocities because of the continuity of mass and enthalpy transport \citep[see][]{RieutordZahn95}. 

Convective downdrafts penetrate into the underlying radiative region before being stopped by its stable stratification that strongly hampers motions in the direction of the local gravity (the vertical or radial direction). A model for such a penetrative convection was proposed by \citet{Zahn91}. It  shows that the downdrafts  enforce a nearly adiabatic stratification over almost  the entire vertical extent of their penetration before being stopped  in a  terminal layer of a few kilometers thickness below which the stratification becomes strongly sub-adiabatic as in the absence of penetrative convection. 
The mean extension of the penetration of the downdrafts depends on a free parameter in Zahn's model and it is usually assumed to be of the order of $(0.1-0.2)\, H_{\rm p}$  at the base of the convection zone, although recent numerical simulations have shown that it depends on the density gradient at the interface between the convection zone and the radiative zone and on the stellar rotation that affects downdrafts through the effect of the Coriolis force \citep[cf.][and references therein]{KorreFeatherstone21}. The statistics of the plume penetration has been analysed by \citet{Prattetal17} and subsequently by \citet{Baraffeetal21} and \citet{Baraffeetal23} by means of numerical simulations of stellar convection. 

The stratification induced by the downdrafts in the overshoot layer is slightly sub-adiabatic $\nabla - \nabla_{\rm ad} <0$\footnote{The temperature gradients are defined as $\nabla_{\rm ad} = (\partial \ln T/\partial \ln P)_{\rm ad}$ at constant entropy and $\nabla = \partial \ln T /\partial \ln P$ in the actual internal stratification \citep[cf.][]{Kippenhahnetal13}.}  \citep[cf. Sect.~5 of][]{Zahn91}. For the typical ranges of adopted parameters, the sub-adiabatic gradient is $|\nabla-\nabla_{\rm ad}| \sim 10^{-6}-10^{-5}$ in Sun-like stars. This  strongly reduces the buoyancy instability of horizontal magnetic fields thus making the overshoot layer a candidate for the storage of the intense magnetic fields (up to $\sim 10$~T) that are invoked by some  models to account for the solar and stellar magnetic activity, in particular, for the formation of sunspots \citep[e.g.][see also Sect.~\ref{model}]{Caligarietal95}.  

The overshoot layer is relevant not only as a potential storage site for intense magnetic fields, but it plays an active and relevant role in the internal stellar dynamics. It is the layer where the downdrafts can excite gravity waves that contribute to the transport of angular momentum in addition to those excited by turbulent convection at the base of the convective envelope \citep[e.g.][]{Zahnetal97, LecoanetQuataert13,Alvanetal14,Pinconetal16,Bretonetal22}. Moreover, penetrative convection contributes to the internal mixing leading to the burning of light elements such as lithium, boron,  and beryllium \citep[e.g.][]{MontalbanSchatzman00}. %\LEt{ This paper is primarily written in UK convention. So, e.g. and i.e. should not have commas.***}

The overshoot layer is regarded as coincident with most of the solar (and stellar) tachocline, namely, the layer where a strong radial gradient of the rotation velocity is localised, at the interface between the rigidly rotating radiative interior and the differentially rotating convection zone \citep{Schouetal98}. The origin of the solar tachocline is still strongly debated with hydrodynamical models essentially relying on the proposal by \citet{SpiegelZahn92}, invoking a strongly anisotropic turbulent viscosity in the layer, while magnetohydrodynamics models mainly rely on dynamo-generated oscillating magnetic fields \citep[cf.][]{Bernabeetal17} since numerical simulations showed the difficulty in confining the shear in the presence of a fossil stationary magnetic field in the radiative zone \citep[e.g.][]{BrunZahn06,Strugareketal11}.  

In the present work, we investigate a possible effect of the equilibrium tide on the convective downdrafts in a late-type MS star that has a close-by massive companion in the case when stellar  rotation is not synchronised with the orbital motion. 
Our investigation has been inspired by the intriguing observational results concerning the  stellar activity of \object{WASP-18} that hosts a planet with a mass of $\sim 10$ Jupiter masses orbiting in only 0.941 days \citep{Hellieretal09}. Such a star was noticed as a member of the group of hosts of massive transiting planets having a very low level of chromospheric activity as measured by the $\log R^{\prime}_{\rm HK}$ index based on the Ca II H\&K resonance line core emission \citep[cf. Fig.~2 of][]{Fossatietal13}. The proposed explanation for such a missing or strongly depressed emission in the core of chromospheric lines  is an absorption by a circumstellar torus of material formed by the strong evaporation of the planetary atmosphere because of their proximity  to their host stars \citep[cf.][]{Haswelletal12,Lanza14,Fossatietal15}. 

A prediction of the absorbing torus model is that the X-ray emission observed in these planetary hosts should be that expected  on the basis of their rotation rate and age. The reason is that the torus is virtually transparent at wavelengths below $\sim 35$~nm, thanks to the proportionality of the bound-free absorption cross section of Hydrogen to the third power of the wavelength. Given the quite fast rotation of WASP-18 ($v\sin i \sim 11$~km~s$^{-1}$, corresponding to a minimum rotation period $P_{\rm rot} \sim 5.6$~days) and an estimated age of $\sim 0.6-0.8$~Gyr \citep{Hellieretal09}, its non-detection at X-ray wavelengths by \citet{Pillitterietal14} came as a big surprise. Their X-ray luminosity upper limit was two orders of magnitude below the expected level making Pillitteri et al. speculate that the massive close-by companion of WASP-18 can somehow hinder its magnetic activity. 
Subsequent observations in the extreme ultraviolet (EUV) with the Hubble Space Telescope by \citet{Fossatietal18} confirmed the suppressed level of magnetic activity also in the EUV emission lines coming from the upper chromosphere and the transition region of the stellar atmosphere, reinforcing its peculiarity and the need for an explanation. 

Recently, \citet{Pillitterietal23}  observed KELT-24, a star very similar to WASP-18 in age and effective temperature, hosting a more distant hot Jupiter, and found it to show the level of X-ray emission predicted on the basis of its young age. A similar level of X-ray emission was observed in other planetary hosts with similar spectral types \citep[cf. Fig.~7 of][]{Pillitterietal23}. Moreover, KELT-24 shows a significantly higher mean energy of the optical flares than WASP-18,  a result based on the observations performed by the Transit Exoplanet Survey Satellite \citep[TESS;][]{Rickeretal16}, thus confirming an unexpected low level of magnetic activity in our  target. 

{A possible explanation for the low level of activity of WASP-18 could be that we have caught the star in a grand minimum of activity such as those affecting the Sun for about 15\% of the time \citep[e.g.][]{Biswasetal23}. Our knowledge of similar phenomena in other stars is extremely limited \citep[cf.][]{Baumetal22,Isaacsonetal24} and no generally accepted physical mechanism has been proposed for the generation of grand minima, even in the Sun \citep{Karak23}. Therefore, we decided to look for an alternative explanation, although the possibility that WASP-18 be  presently in a grand minimum regime cannot be excluded.}

Specifically, we propose here a conjectural model that predicts the disruption of the downdrafts because of the turbulence excited below the convection zone by the motions associated with the equilibrium tide. This leads to a significant increase of the sub-adiabaticity in the layers immediately below the stellar convection zone that can make the buoyancy instability of magnetic fields stored there much more difficult. This  hinders the formation of starspots in the stellar photosphere resulting in a strong reduction of stellar magnetic activity and offering a potential explanation for the very low level of activity observed in WASP-18.

%--------------------------------------------------------------------
\section{Model}
\label{model}

{\subsection{Overshoot layer and  hydromagnetic dynamos in late-type stars}
\label{dynamo_models}

The overshoot layer has been proposed as a seat for the storage of large amount of magnetic flux in late-type stars. Even if the main stellar hydromagnetic dynamo works in the bulk of the convection zone as suggested by recent numerical simulations \citep[e.g.][]{BrunBrowning17,Kapylaetal23} and the overshoot layer may not play a universal role in the amplification and modulation of the stellar magnetic field as discussed, for instance, in Sect.~8.4 of \citet{Charbonneau20}, we speculate that the overshoot layer  plays a relevant role for the dynamo action in the case of F-type stars such as WASP-18. 

The larger velocities of convective motions in F-type stars with respect to G- and K-type stars \citep{Brunetal17,Corsaroetal21} can lead to the formation of stronger downdrafts as seen in the numerical simulations of, for example,  \cite{Bretonetal22}. These can effectively pump most of the magnetic flux produced by a dynamo from the bulk of the convection zone into the overshoot layer  \citep[e.g.][]{Tobiasetal98,Tobiasetal01} that we assume to coincide with the stellar tachocline. Thanks to the  tachocline shear, the magnetic field in the overshoot layer can be remarkably amplified, thus contributing most of  the magnetic flux responsible for stellar activity, in particular, for photospheric starspots. 

 An important finding is that a dynamo model with an overshoot layer is capable of producing a more stable modulation with longer cycles than a dynamo model restricted to the convection zone \citep[][]{Guerreroetal16,Kapylaetal23}. This seems to be in closer agreement with the persistent decadal activity cycles observed in stars with a radiative interior in contrast to the more chaotic activity displayed by fully convective stars.  %In other words, if the flux stored inside the overshoot layer is not capable to emerge to the stellar photosphere, the magnetic activity level of an F-type star is strongly reduced.

Another class of dynamo models that could be relevant for WASP-18 is that proposed to operate close to the top of the convection zone, precisely in a near surface shear layer \citep{BohmVitense07,Brandenburgetal17}  akin to that found by helioseismology in the Sun \citep{Schouetal98}.  Assuming that such a near-surface dynamo is operating in WASP-18, we predicted a negligible effect of the tidal flow on it because the tidal shear is at least one to two orders of magnitude smaller than the radial shear, assuming the latter to be similar to that observed in the Sun \citep[see Sect.~\ref{interaction} for a quantification of the tidal shear and, for the near surface shear,][]{Barekatetal14}. Moreover, the curl of the equilibrium tidal flow vanishes in the convection zone \citep{Ogilvie14}, thereby preventing the equilibrium  tide from affecting the $\alpha$ effect of the dynamo in the convection zone. In other words, we do not expect tides to be capable of significantly affect or quench a near surface dynamo, when it is indeed at work.
Similarly, we do not expect that the equilibrium tide can produce any relevant perturbation of the shaping action of the sub-surface shear layer on the magnetic fields generated in the bulk of the convection zone \citep[][]{PipinKosovichev11,Brandenburgetal23}. 

Nevertheless, a near-surface dynamo is likely not capable of amplifying the magnetic field  as effectively as a dynamo  storing the toroidal field in the overshoot layer because the strong superadiabaticity in near surface layers makes a comparable toroidal field strongly unstable to buoyancy there. \textcolor{black}{Moreover, the observed solar meridional flow is poleward in the upper half of the solar convection zone, thus leading to a reduced or reversed migration of the active regions with respect to the observed butterfly diagram in the Sun, if the toroidal field generation is dominated by a near-surface solar dynamo \citep[see e.g.  ][for the role played by the helioseismically determined meridional circulation in a Babcock-Leighton model of the solar cycle]{Cloutieretal23}.} For such reasons, we focus our investigation on dynamos where the overshoot layer below the convection zone plays a crucial role contributing most of the magnetic flux responsible for stellar activity. \textcolor{black}{Such dynamos have the advantage of storing strong toroidal magnetic fields ($\geq 10$~T) in the subadiabatic overshoot layer. These fields can emerge at low latitudes in the Sun producing the active regions observed during the maximum and decay phase of the solar cycle. On the other hand, dynamos storing the toroidal fields in the deepest layers of the superadiabatic convection zone, such as Babcock-Leighton dynamos \citep{Charbonneau20,Cloutieretal23}, are generally limited to weaker fields ($\sim 10^{4}$~G), that lead to active regions formation only at high latitudes ($\geq 30^{\circ}-40^{\circ}$) in the Sun \citep[cf.][]{Caligarietal95}. }
}

\subsection{Downdrafts and sub-adiabatic stratification in the overshoot layer}
\label{downdraft_parameters}

The penetrative convection below the convective zone has been described as consisting of downward turbulent plumes (or downdrafts) by \citet{RieutordZahn95}. Their model was extended in Sect.~3 of \citet{Pinconetal16} and we adopt it here to estimate the relevant parameters of the convective downdrafts. Considering an horizontal section of a plume, its downward directed mean velocity is expressed by \citet{Pinconetal16} by a Gaussian profile with a standard deviation (effective radius of the downdraft): 
\begin{equation}
    b = \frac{z_{0}}{\sqrt{2}} \frac{3 \alpha_{\rm E}(\Gamma_{1}- 1)}{2\Gamma_{1}-1}, 
    \label{downdraft_rad}
\end{equation}
where $z_{0}$ is the depth of the convection zone, $\alpha_{\rm E} =0.083$ is an universal coefficient derived from laboratory studies and simulations of penetrative convection \citep[cf. Sect.~3.2 of][]{Pinconetal16}, and $\Gamma_{1}$ is the ratio of the specific heats at constant pressure and volume, that is, $\Gamma_{1} = 5/3$ in the fully ionised layers in the interior of a solar-like star. The typical value of the downdraft cross-sectional size $b$  is between $5\times 10^{6}$ and $10^{7}$~m in late-type MS  stars. The total number of plumes ${\cal N}$ simultaneously present at the base of the convection zone is of the order of $\sim 10^{3}$ according to \citet{RieutordZahn95},  which implies an areal filling factor $f \sim 0.1$. 

The downward directed plume velocity has been estimated by assuming that the downdrafts transport all the stellar convective flux as in Sect.~3 of \citet{RieutordZahn95}. In such a way, their velocity at the top of the overshoot layer is given by \citep[cf. Eq.~(25) of][]{Pinconetal16}:
\begin{equation}
    V_{\rm b} = \left( \frac{2L}{\pi \rho_{\rm b} r_{\rm b}^{2}} \right)^{1/3},
    \label{downdraft_vel}
\end{equation}
where $L$ is the luminosity of the star, $\rho_{\rm b}$ the density at the base of the convection zone, and $r_{\rm b}=R-z_{0}$ the radius of that layer with $R$ the radius of the star. Adopting the model by \citet{Zahn91}, the plume velocity depends on the depth $z$ as measured from the level $r=r_{\rm b}$ according to
\begin{equation}
    V(z) = V_{\rm b} \left[ 1- \left( \frac{z}{L_{\rm p}}\right)^{2} \right]^{1/3}, 
    \label{downdraft_vel1}
\end{equation}
where $L_{\rm p}$ is the depth down to which the plume penetrates, that is, the depth of the overshoot layer. Here it is assumed to be $0.1\,H_{\rm p}$, where $H_{\rm p}$ is the pressure scale height at the base of the convection zone \citep[see][for a discussion of  its dependence on stellar parameters including the effect of rotation]{KorreFeatherstone21}. {However, the possibility of a  deeper convective penetration, down to $\sim 0.5\, H_{\rm p}$, cannot be excluded according to the results by \citet{Andersetal22} who discussed the difficulties for a proper numerical simulation of overshooting convection in stars. }

The mean lifetime of the convective downdrafts  $\tau_{\rm p}$ is the most uncertain parameter. Nevertheless, following the arguments in Sect.~3.4 of \citet{Pinconetal16}, it seems reasonable to assume it of the order of the convective turnover time at the base of the stellar convection zone, that is, 
\begin{equation}
\tau_{\rm p} \sim \alpha_{\rm mlt} H_{\rm p}/\varv_{\rm c},
\label{tau_downdraft}
\end{equation}
where $\alpha_{\rm mlt}$ is the ratio of the mixing length to the local pressure scale height $H_{\rm p}$, while $\varv_{\rm c}$ is the velocity of convective motions at the base of the convection zone as given by the standard mixing-length theory \citep[cf. Ch.~7 of][]{Kippenhahnetal13}. 

The time evolution of the downdrafts is uncertain as well. It has been discussed by \citet{Pinconetal21} who considered two limiting cases where their velocity field varies as $\exp(-t^{2}/\tau_{\rm p}^{2})$ or $\exp (-t/\tau_{\rm p})$, respectively. The former Gaussian time dependence corresponds to a rapid damping of the downdrafts, while the latter exponential implies a slower decay and a longer duration. We  adopted the exponential decay law because we are interested in the case of relatively long-lived downdrafts. Moreover, this assumption  allows us to  easily add the effect of the turbulent diffusivity on their evolution (cf. Sect.~\ref{modification}). 

In conclusion, we consider the following vertical velocity field for our downdrafts:
\begin{equation}
    {\vec \varv}_{\rm down} (s, z, t) = V(z) \exp{\left[- \left( \frac{t}{\tau_{\rm p}}\right) \right]} \exp \left[ -\left( \frac{s^{2}}{2b^{2}} \right) \right] \hat{\vec z},
\label{downdraft_vfield}
\end{equation}
where $s$ is the horizontal distance from the axis of the downdraft assumed to be cylindrically symmetric around its vertical axis \citep[cf.][]{RieutordZahn95}, $\hat{\vec z}$ is the unit vector in the vertical direction, $V(z)$ is given by Eq.~\eqref{downdraft_vel1},  $b$ is the effective radius of the downdraft as given by Eq.~\eqref{downdraft_rad}, and the time, $t,$ is measured from the starting of the downdraft.  

The mean temperature gradient in the overshoot layer down to the level $z \sim L_{\rm p}$ is given by \citep[cf. Eq.~(5.10) in ][]{Zahn91}
\begin{equation}
    \frac{\nabla}{\nabla_{\rm ad} - \nabla} = \frac{f}{c\chi_{\rm p}} \frac{V_{\rm b} H_{\rm p}}{K}, 
    \label{subadiabaticity}
\end{equation}
where $f$ is the filling factor of the downdrafts at the base of the convection zone where $r=r_{\rm b}$, $c \sim 1- f$ a geometric coefficient, $\chi_{\rm p} = (\partial \ln \chi / \partial \ln p )_{\rm ad}$ is the logarithm derivative of the radiative conductivity, $\chi,$ with respect to the plasma pressure $p$ in adiabatic conditions, and $K = \chi/(\rho \, c_{\rm p})$ is the thermal diffusivity with $\rho$ as the density and $c_{\rm p}$ the specific heat at constant pressure. The radiative conductivity depends on the opacity of the plasma $\kappa$ as
\begin{equation}
    \chi = \frac{16\, \sigma \, T^{3}}{3\, \rho \, \kappa},
\end{equation}
where $\sigma$ is the Stefan-Boltzmann constant and $T$ the temperature. 
In the case of the Sun, $\chi_{\rm p} \sim 1.8$, $K \sim 2 \times 10^{3}$~m$^{2}$~s$^{-1}$ \citep{Zahn91}, $V_{\rm b} \sim 190$~m~s$^{-1}$ \citep{Pinconetal16}, $H_{\rm p} \sim 5.6 \times 10^{7}$~m, so that ${V_{\rm b} / H_{\rm p}}{K \sim 3.5 \times 10^{6}}$. 
For $ \nabla \simeq \nabla_{\rm ad} = 0.4 $ and $f \sim 0.1$, Eq.~\eqref{subadiabaticity} gives a sub-adiabaticity $\nabla_{\rm ad} - \nabla \sim 2.5 \times 10^{-5}$.

\subsection{Interaction of the equilibrium tide with the downdrafts }
\label{interaction}

The equilibrium tide is the quasi-hydrostatic deformation of the host star under the action of the tidal potential generated by its planet \citep{Remusetal12,Ogilvie14,Barker20}. The relative deformation in the radial direction is of the order of 
\begin{equation}
    \epsilon (r) \equiv \left( \frac{m_{\rm p}}{M}\right) \left( \frac{r}{a}\right)^{3},
\end{equation}
where $m_{\rm p}$ is the mass of the planet, $M$ the mass of the star, $r$ the radial distance from the centre of the star  where we consider the tidal deformation, and $a$ the semimajor axis of the orbit assumed to be circular and in the equatorial plane of the star. The tidal deformation $\vec \xi$ in the radiative zone of a star is a solenoidal vector ($\nabla \cdot {\vec \xi} = 0$) as shown by \citet{Ogilvie14} in his Sect.~3.2. 

The vertical deformation due to the equilibrium tide at a distance $r$ from the centre of the star is $\xi_{\rm r} \sim \epsilon \, r$, while the  deformation in the azimuthal (horizontal) direction $\xi_{\phi}$ follows from $\nabla \cdot {\vec \xi}=0$  by considering that the horizontal wavenumber of the tidal deformation is $2/r$ for the semi-diurnal tide at the equator. In other words, we have 
\begin{equation}
\xi_{\phi} \sim 2 \, \epsilon \, r.
\label{xi_phi}
\end{equation}
The Eulerian velocity field of the equilibrium tide in a reference frame that rotates with the stellar angular velocity $\Omega$ is $\partial{\vec \xi} /\partial t = \omega_{\rm tide}\, {\vec \xi}$, where $\omega_{\rm tide}$ is the frequency of the semi-diurnal tide given by
\begin{equation}
    \omega_{\rm tide} = 2 (n - \Omega), 
\end{equation}
where $n \equiv 2\pi/P_{\rm orb}$ is the orbital mean motion with $P_{\rm orb}$ being the orbital period and $\Omega = 2\pi/P_{\rm rot}$ the angular velocity of rotation of the star with a rotation period $P_{\rm rot}$, and we assume $n> \Omega$ for a close-by planet. We consider only the semi-diurnal or quadrupolar equilibrium tide (with spherical degree and azimuthal wavenumbers $l=m=2$ in Remus's or Ogivie's notation) because the higher-order tidal terms are much smaller and can be neglected for our purposes.

%We assume that each downdraft extends from near the top of the stellar convection zone where $r \sim R$ down to the base of the convection zone at $r \sim r_{\rm b}$ and in the overshoot layer below. {\em We regard each downdraft as a rigid vertical structure that is displaced horizontally by an amount equal to $\xi_{\phi}(R)$ by the equilibrium tide, while the surrounding plasma is displaced in proportion to the distance $r$ from the center of the star. Adopting this assumption, a relative displacement $\xi_{\phi}(r)-\xi_{\phi}(R)$ is produced at a distance $r$ from the center of the star.} Specifically, at the depth $r= r_{\rm b}$ corresponding to the base of the convection zone, the relative displacement between a downdraft and the surrounding plasma is 
%\begin{equation}
%\Delta \xi_{\phi} \sim 2 \epsilon_{\rm b} \, (R-r_{\rm b}), \mbox{ with %$\epsilon_{\rm b} = \epsilon(r_{\rm b})$,}
%\label{delta_xi_phi}
%\end{equation}
%while the relative velocity between the downdraft and the tidal flow there is
%\begin{equation}
%\Delta \dot{\xi}_{\phi} \sim \omega_{\rm tide} \, \Delta \xi_{\phi} \sim \omega_{\rm tide} \epsilon_{\rm b} \, (R - r_{\rm b}) \sim 4 \epsilon_{\rm b} (n -\Omega) (R - r_{\rm b}).
%\end{equation}

In the case of the massive very hot Jupiter  of WASP-18, with $a = 0.02014$~au, $m_{\rm p}=10.06$ Jupiter masses, $M=1.22$~M$_{\odot}$, $P_{\rm rot} \sim 5.6$~days,  $P_{\rm orb}=0.941$~days, and $r_{\rm b}/R = 0.844$ with $R=1.21$~R$_{\odot}$ being the star radius (cf. Sect.~\ref{stellar_model}), we obtain $\epsilon(r_{\rm b}) \equiv \epsilon_{\rm b} = 1.0 \times 10^{-4}$ and $\omega_{\rm tide} = 1.28 \times 10^{-4}$~s$^{-1}$. Hence, the azimuthal relative tidal velocity between a downdraft and the surrounding plasma in the overshoot layer turns out to be  $ \dot{\xi}_{\phi} \sim 18$~m~s$^{-1}$. 

The shear associated with the equilibrium tide is too small to produce a linear instability leading to turbulence \citep[][Sect.~4e]{Seguin76,Zahn77}. Therefore, the tidal flow is generally assumed to be laminar, except in the cases when the tide enters into a parametric resonance  with inertial waves  \citep[cf. Sect.~4.1 of][the so-called elliptic instability]{Ogilvie14}  or gravity waves in the deep radiative interior \citep{Vidaletal19, Weinbergetal12}. If such a resonance occurs, the resulting nonlinear instability can lead to a turbulent flow that increases the dissipation by many  orders of magnitude. However, none of these situations occurs in the system of WASP-18. Specifically, the elliptic instability requires $-1 < n/\Omega < 3$, a condition that is not verified in this system where $n/\Omega \sim 5.5$. On the other hand, $n/\Omega \sim 5.5$ could allow a resonance with gravity waves in the deep radiative interior where the Brunt-V\"ais\"ala frequency $N \ga (8-10) \, \Omega$ as illustrated in Fig.~2 of \citet{Vidaletal19}. Although the Brunt-V\"ais\"ala frequency $N$ increases very rapidly with depth below the top of the radiative zone, the layers where such a resonance can occur are probably located much deeper than the overshoot layer; thus, the resulting turbulence cannot affect the layer we are interested in. {Similarly, any resonance with inertial waves can be excluded in WASP-18 because those waves cannot be excited in the interior of our target given that the tidal frequency is greater than twice its angular velocity of rotation \citep{OgilvieLin07}.}

{We conjecture that} the establishment of a turbulent flow in the overshoot layer can be related to a different mechanism that depends on the interaction between the vertical convective plumes that straddle the layer and the horizontal tidal flow in the azimuthal direction that is orthogonal to them. We assume that the mean velocity field of the downdrafts is not modified by the presence of the equilibrium tide. We regard each downdraft as a rigid structure, that is, a vertical obstacle to the horizontal flow of the equilibrium tide. The rigidity of the downdrafts is due to the presence of a relatively strong vertical magnetic field in the downdrafts as seen in numerical simulations of penetrative convection and magnetic flux pumping into the overshoot layer \citep[see e.g. Fig.~2 of][]{Tobiasetal98}. Such a field is produced by the amplification, owing to the vertical shear in a downdraft, of a weaker field in the overlying convection zone that is provided by an hydromagnetic dynamo operating there. Since the downdraft velocity is of the order of hundreds of meters per second   (see Sects.~\ref{downdraft_parameters} and~\ref{stellar_model}), the vertical magnetic field in a downdraft has an energy density about two orders of magnitude larger than the kinetic energy density of the horizontal tidal flow $\dot{\xi}_{\phi}$ that is only of the order of tens of meters per second. Therefore, such a vertical magnetic field cannot be significantly distorted by the tidal flow and the magnetised plasma of the downdraft provides an obstacle to the tidal flow. 

\begin{figure}
%   \centering
\hspace{-0.3cm}
   \includegraphics[width=10cm,height=8cm,angle=0]{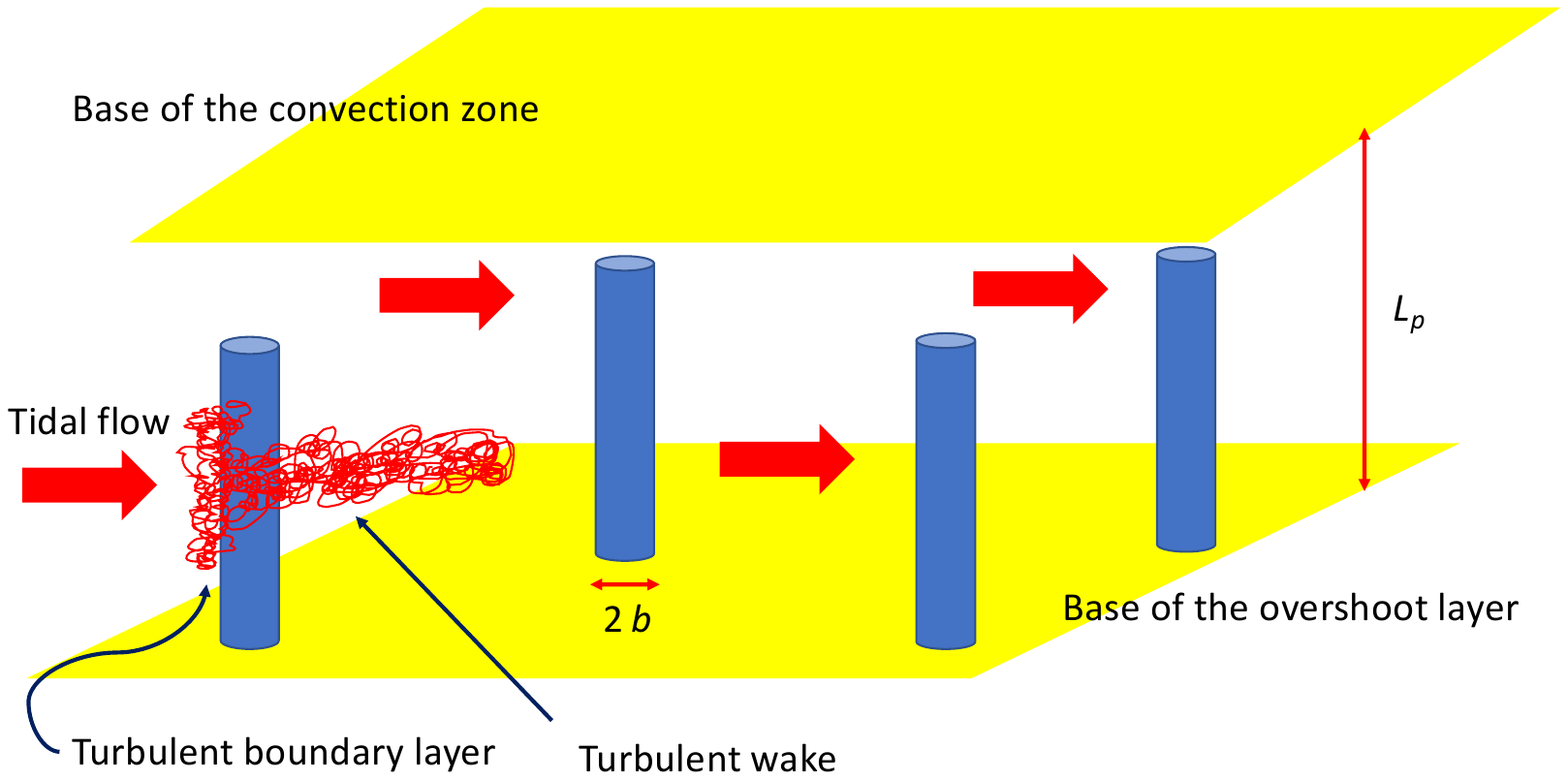}
   \vspace{-2.0cm}
   \caption{Cartoon showing the convective downdrafts in the overshoot layer and the turbulence produced by the horizontal tidal flow moving past them in the layer. }
   \label{fig_turb}
    \end{figure}

In Fig.~\ref{fig_turb}, we provide a sketch of our system with the magnetised plumes shown as blue rigid vertical cylinders of diameter $2b$, while the horizontal yellow planes indicate the base of the convection zone (top) and the base of the overshoot layer at a depth $L_{\rm p}$ below (bottom), respectively. The horizontal red arrows indicate the horizontal flow of the equilibrium tide, while turbulence is sketched as knotted red solid lines around and past only part of one of the cylinders,   for simplicity. Both the turbulent boundary layer and the turbulent wake are schematically indicated. We note that although turbulent flow lines have been drawn for  only a small part of the domain, the turbulence extends itself over the whole overshoot layer because of the very high Reynolds number of such a system (see Sect.~\ref{oscillating_flow}).

{An additional process occurring in  the overshoot layer is the formation of a more or less uniform horizontal layer of magnetic field owing to the pumping effect of the downdrafts. We expect such a magnetised layer to be located in the lower part of the overshoot domain, not far from the terminal layer where the downdrafts are braked by the stratification of the underlying radiative zone. This is because the downdrafts themselves are capable of pushing the field down to that level. In other words, we assume that the upper part of the overshoot layer is devoid of any horizontal magnetic field and possesses  only the vertical fields associated with the downdrafts.  Therefore, in the upper part of the overshoot layer, the turbulence around the downdrafts cannot be quenched by the horizontal magnetic field. }

\subsection{Stationary and oscillating horizontal flows}
\label{oscillating_flow}

It is useful to consider an analogous system consisting of an array of vertical rigid indefinite circular cylinders placed in a flow perpendicular to their axes of symmetry {that is produced by the equilibrium tide}. {Let us first consider the case of a uniform and steady perpendicular flow.} In this system, the Reynolds number is defined as $Re = D U /\nu$, where $D$ is the diameter of each cylinder, $U$ the speed of the flow at infinity, and $\nu$ the kinematic viscosity of the fluid. The kinematic microscopic viscosity below the solar convection zone is $\nu \sim 3 \times 10^{-3}$~m$^{2}$~s$^{-1}$ \citep[cf. Table 1 of][]{BrunZahn06}; assuming $D \sim 2 b \sim 10^{7}$~m and $U \sim  \dot{\xi}_{\phi} \sim 18$~m~s$^{-1}$, we obtain $Re \sim 6 \times 10^{10}$ that corresponds to a hydrodynamic regime with a strongly developed turbulence. For comparison, the appearance of turbulence in the wake of the flow past a single indefinite cylinder occurs for $Re \ga 400$, while the turbulence fully develops in the boundary layer at the surface of the cylinder for $Re \ga 5 \times 10^{5}$ \citep[e.g.][]{Zdravkovich97}. {\citet{Williamson96} reported that, although for $Re > 10^{5}-10^{7}$ a uniform fully turbulent wake is expected, a periodic turbulent vortex shedding is observed as typically found in  cylinder wakes at $Re \sim 10^{3}-10^{4}$. The vortex shedding frequency $\omega_{\rm vs}$ enters into the definition of the non-dimensional Strouhal number $S_{\rm n} \equiv \omega_{\rm vs} D/U$ that reaches an asymptotic value of $\sim 0.20-0.25$ with increasing Reynolds number $Re$. 

When the flow orthogonal to the axis of a cylinder is oscillating as in the case of the equilibrium tidal flow, a significant change in the above stationary picture occurs when the frequency of the flow oscillation or its first harmonic come close to the natural shedding frequency $\omega_{\rm sv}$ corresponding to the asymptotic value of the Strouhal number. In such a case, the shedding frequency tends to lock to the oscillation frequency of the flow and the coherence lengthscale along the wake can increase remarkably \citep[cf. ][]{Bearman84,Kodkharetal21}. However, such a resonant regime is not approached in the case of WASP-18 because the natural shedding frequency for a cylinder of radius $\sim 6.6 \times 10^{6}$~m in a flow of velocity $U \sim 18$~m~s$^{-1}$ is $\omega_{\rm vs} \sim 3 \times 10^{-7}$ s$^{-1}$ for an asymptotic  Strouhal number $S_{\rm n} = 0.25$, while the tidal frequency is higher by three orders of magnitude being $\omega_{\rm tide} \sim 1.3 \times 10^{-4}$~s$^{-1}$. Therefore, we do not expect any significant modification in the cylinder wake due to the periodic oscillations of the tidal flow.  }

{In the present model, the motivation for treating the downdrafts as obstacles to the tidal flow is the presence of a vertical magnetic field in the downdrafts themselves with an intensity larger than the equipartition strength corresponding to the tidal flow. However, it is interesting to investigate the interaction between the tidal flow and the downdrafts also adopting a different viewpoint that does not rely on the presence of a strong vertical magnetic field in the downdrafts. %In the absence of such a field, we should adopt a different hypothesis to describe the interaction between a downdraft and an oscillating tidal flow. 

Specifically, we consider the linear modelling of the interaction between a columnar vortex and an incoming wave  by \citet{Dandoyetal23}. Convective downdrafts in  rotating stars are places of vorticity amplification due to the vertically directed velocity of the plasma that increases with depth in the overlying convection zone. Therefore, we expect that our downdrafts behave similarly to vortices in the overshoot layer. An incident planar wave excites oscillations of a vortex with its characteristic frequency that lead to the radiation of waves from the vortex. Those waves have cylindrical symmetry in the far field and  carry between 10\% and 20\% of the energy of the incident wave depending on whether resonant or neutral oscillation modes are excited in the vortex. 

In our case, each downdraft becomes a source of cylindrical waves that can interact nonlinearly with each other producing a turbulent energy cascade \citep[e.g.][]{Nazarenko11}. The turbulent viscosity produced by such a process should be evaluated by a nonlinear model of the wave excitation and interaction; however, it should be comparable within one order of magnitude with our estimate (given in Sect.~\ref{turbulence_model}). This is based on the fact that the energy of the interacting waves can reach $0.1-0.2$  times the energy of the incident tidal wave, while their wavelength is at least comparable with the mean radius of the vortices \citep{Dandoyetal23}. }

\subsection{Turbulence in the overshoot layer}
\label{turbulence_model}

The sub-adiabatic stratification of the overshoot layer has a remarkable impact on the turbulence and the turbulent diffusivity that {we expect to} develop inside it as a consequence of the interaction between the downward directed convective plumes and the equilibrium tidal flow. The stable stratification strongly hampers motions in the direction of the local gravity, while the motions over planes perpendicular to the gravity are constrained by the Coriolis force \citep[e.g.][]{Mathisetal18}. This leads to an anisotropic turbulent viscosity whose value in the horizontal direction is larger than in the vertical direction. 

The turbulent diffusivity in the horizontal direction can be written as 
\begin{equation}
\nu_{\rm turb \, h} \sim \ell u^{\prime},
\label{hor_diff}
\end{equation}
where $\ell$ is the lengthscale of the horizontal turbulence, that is, in the plane perpendicular to the local acceleration of gravity, while $u^{\prime}$ is the turbulent velocity in the same horizontal plane that comes from the interaction between the vertical downdrafts and the tidal horizontal flow. %Therefore, in principle, $u^{\prime}$ can be different from the horizontal tidal velocity. 
In the next subsection, we make educated guesses on $\ell$ %those two basic quantities 
taking into account the effect of the Coriolis force that limits the horizontal size of the turbulent eddies. After having estimated the horizontal component of the turbulent diffusivity, we estimate its vertical counterpart $\nu_{\rm turb \, v}$ by means of the anisotropic turbulence model by \citet{Mathisetal18} that generalises a previous quasi-linear model by \citet{KitchatinovBrandenburg12}. 

\subsection{Length scale of the horizontal turbulent motions}
\label{length_scale}

The velocity field $\vec u$ in the overshoot layer outside the convective plumes  can be written as
\begin{equation}
    {\vec u} = {\vec u}_{0} + {\vec u}^{\prime},
    \label{vel_def}
\end{equation}
where ${\vec u}_{0}$ is the tidal velocity field and ${\vec u}^{\prime}$  the turbulent velocity field produced by the interaction of the tidal flow with the penetrating plumes because of its very large Reynolds number (cf. Sects.~\ref{interaction} and~\ref{oscillating_flow}). 
We approximate the turbulent velocity field as a 2D field in the local horizontal plane, that is,  the plane orthogonal to the local gravity. 

In order to estimate the length scale of this horizontal turbulence, we start from the momentum equation in a rotating reference frame 
\begin{equation}
\frac{\partial {\vec u}}{\partial t} + \left({\vec u} \cdot \nabla \right) {\vec u} = - \frac{1}{\rho} \nabla p + 2 {\vec \Omega} \times {\vec u} + \nabla \Phi + \nabla \Psi, 
\label{mom_eq}
\end{equation}
where $t$ is the time, $\rho$ the plasma density, $p$ the pressure, $\vec \Omega$ the angular velocity of rotation of the reference frame,  $\Phi$ the gravitational potential of the star, and $\Psi$ the tidal potential \citep[cf.][\S~2.1]{Ogilvie14}. For simplicity, we assume that the star is rotating with a uniform angular velocity $\vec \Omega$,   neglect the viscous force because of the very large Reynolds number, and the Lorentz force by assuming that the dynamical effects of the magnetic fields are small outside  the downdrafts {(cf. the final paragraph of Sect.~\ref{interaction})}. 

To treat the turbulent velocity field, we introduce an ensemble mean indicated with  angular brackets and assume that $\langle {\vec u}_{0} \rangle = {\vec u}_{0}$ and $\langle {\vec u}^{\prime} \rangle = 0 $, so that $\langle {\vec u}  \rangle = {\vec u}_{0} $. Moreover, we assume that the ensemble mean commutes with the time derivative, that implies  $ \langle \partial {\vec u}^{\prime}/\partial t \rangle = \partial \langle {\vec u}^{\prime} \rangle /\partial t = 0 $, and with the spatial derivatives.  

The equilibrium tidal flow ${\vec u}_{0}$ is not an exact solution of the momentum equation, but it is an approximation computed by assuming that the star is capable of adjusting its stratification to the time-varying tidal potential with a very small lag, that is, with a small deviation from the hydrostatic equilibrium \citep[e.g.][]{Ogilvie14,Barker20}.  In other words, we may write an equation for the equilibrium tidal flow in the form 
\begin{equation}
\frac{\partial {\vec u}_{0}}{\partial t} + \left({\vec u}_{0} \cdot \nabla \right) {\vec u}_{0} = - \frac{1}{\rho} \nabla p^{\prime} + \nabla \Phi^{\prime} + \nabla \Psi,
\label{tide_eq}
\end{equation}
where the prime indicates the perturbation from the hydrostatic equilibrium we have in the absence of tides and that can be written as  $-\nabla p + \rho \nabla \Phi = 0$ \citep[cf.][\S~3.2]{Ogilvie14}. In Eq.~\eqref{tide_eq}, the Coriolis force has been neglected which is possible because the tidal frequency $\omega_{\rm tide}$ is remarkably larger than the stellar spin frequency $\Omega$ given that we are considering planets orbiting very close to slowly rotating late-type stars. Moreover, we neglected the density perturbation by assuming that the tidal displacement ${\vec u}_{0} \,\omega_{\rm tide}^{-1}$ is small in comparison with the density scale height in the overshoot layer and that the tidal velocity is much smaller than the local sound speed, so that $\nabla \cdot {\vec u}_{0}=0$. 

Subtracting Eq.~\eqref{tide_eq} from Eq.~\eqref{mom_eq}, taking the component in the meridional direction ${\vec e}_{\theta}$ so that ${\vec e}_{\theta} \cdot \nabla \Phi = 0$, and applying the ensemble mean to average out some of the terms containing the turbulent velocity field, we find
\begin{equation}
\langle \left({\vec u}^{\prime} \cdot \nabla \right) {\vec u}^{\prime} \rangle \cdot {\vec e}_{\theta} = 2 \left({\vec \Omega} \times {\vec u}_{0} \right) \cdot {\vec e}_{\theta},
\label{theta_comp}
\end{equation}
that shows how the spatial correlations of the components of the turbulent velocity are ruled by the Coriolis force acting on the mean tidal flow. We note that the Coriolis force has been retained in the momentum equation for the turbulent velocity because the characteristic turbulence time scale may be much longer than the rotation period of the star. 

The scalar triple product on the right-hand side of Eq.~\eqref{theta_comp} can be recast as 
$2 \left({\vec \Omega} \times {\vec u}_{0} \right) \cdot {\vec e}_{\theta} = -2 {\vec \Omega} \cdot \left({\vec u}_{0} \times {\vec e}_{\theta} \right) = -2 \Omega \,  u_{0\, \phi}\cos \theta $, where $\theta$ is the colatitude measured from the North pole and $u_{0 \,\phi}$ is the component of the tidal flow in the azimuthal direction. In such a way, Eq.~\eqref{theta_comp} becomes
\begin{equation}
\langle \left({\vec u}^{\prime} \cdot \nabla \right) {\vec u}^{\prime} \rangle \cdot {\vec e}_{\theta} = -2 \Omega u_{0\, \phi} \cos \theta. 
\label{theta_comp1}
\end{equation}

The characteristic length scale $\ell $ of the turbulent motions  in the horizontal plane can be estimated from Eq.~\eqref{theta_comp1} by considering that the largest turbulent eddies have a velocity amplitude $u^{\prime}$ that is comparable with that of the tidal velocity flow $u_{0}$, so that, in order of magnitude, 
\begin{equation}
\frac{u^{\prime\, 2}}{\ell} \sim 2 \Omega \, u_{0} \cos \theta, \mbox{with  $u^{\prime} \la  u_{0}$.}
\label{om_turb}
\end{equation}
Considering turbulent motions close to the stellar equator, we find $ \theta \sim [(\pi/2) - \ell/r]$, where $r$ is the radius of the spherical surface over which we are estimating the turbulent length scale. Therefore,  Eq.~\eqref{om_turb} becomes
\begin{equation}
\frac{u^{\prime\, 2}}{\ell} \sim 2 \Omega \, u_{0} \frac{\ell}{r},
\label{om_turb1}
\end{equation}
and assuming that $u^{\prime} \sim u_{0} \sim \omega_{\rm tide} \, \xi_{\phi} $, we finally obtain
\begin{equation}
\ell \sim \left[ \left( \frac{\omega_{\rm tide}}{2\Omega} \right) \, r \,  \xi_{\phi} \right]^{1/2}.
\label{ell_eq}
\end{equation}
Given that $ \xi_{\phi} \sim 2 \epsilon_{\rm b} r_{\rm b} \ll r$, the length scale $\ell \ll r$, that justifies the approximation $\cos \theta = \sin (\ell/r) \sim \ell/r$ adopted in Eq.~\eqref{om_turb1}. 

\subsection{Turbulent diffusivities and downdraft diffusion timescale}
\label{turbulent_diff}

By applying the results of the previous Sect.~\ref{length_scale}, the length scale and the typical velocity of the turbulent motions at  $r=r_{\rm b}$ can be estimated. By means of Eq.~\eqref{hor_diff}, we obtain an estimate for the horizontal turbulent diffusivity in the overshoot layer as
\begin{eqnarray}
    \nu_{\rm turb \, h}  & \sim & \omega_{\rm tide} \,  \xi_{\phi} \left[ \left( \frac{\omega_{\rm tide}}{2 \Omega } \right) r_{\rm b} \,  \xi_{\phi} \right] ^{1/2} = \nonumber \\ 
   & = & 2 \Omega^{-1/2} \omega_{\rm tide}^{3/2} \epsilon_{\rm b}^{3/2} r_{\rm b}^{2},
\label{hor_turb_1}
\end{eqnarray}
where the horizontal tidal displacement was substituted from  Eq.~\eqref{xi_phi}. The vertical turbulent diffusivity can be computed from the horizontal diffusivity by applying the model by \citet{Mathisetal18}. Specifically, the ratio of the two diffusivities is given by their Eq.~(41) as
\begin{equation}
    \frac{\nu_{\rm turb \, v}}{\nu_{\rm turb \, h}} =  \frac{2 \Omega^{2}}{N^{4} \tau^{2}},
    \label{mathis_ratio}
\end{equation}
where $\Omega$ is the star rotation frequency, $N$ the Brunt-V\"ais\"ala frequency, and $\tau$ a characteristic time scale of the turbulence. The Brunt-V\"ais\"ala frequency is given by
\begin{equation}
    N^{2} = \frac{g}{H_{\rm p}} \left( \nabla_{\rm ad} - \nabla \right),
    \label{bvf}
\end{equation}
where $g$ is the local acceleration of gravity and the other symbols have already been  introduced above. Equation~\eqref{bvf} is valid when the gradient of the  molecular weight can be neglected and the plasma behaves as an ideal gas. Below the convection zone, $N^{2} >0$, indicating a stable stratification where the buoyancy force opposes adiabatic radial displacements of the plasma elements.  

The ratio between the two diffusivities as given by  Eq.~\eqref{mathis_ratio} depends on  the relative intensities of the two fundamental forces that constrain the turbulent motions in the vertical and the horizontal directions in the stably stratified layers of a rotating star, that is, the buoyancy force, parameterised by the Brunt-V\"ais\"ala frequency, and the Coriolis force, parameterised by the stellar rotation frequency, respectively. 
The characteristic turbulent timescale $\tau$ can be identified with the timescale of variation of the velocity of the largest eddies, that is, $\tau \sim \ell/u^{\prime} \sim \ell/(\omega_{\rm tide} \,  \xi_{\phi}$), where we have approximated the velocity of the largest eddies with the tidal flow velocity. From Eq.~\eqref{ell_eq}, we see that  $\ell/ \xi_{\phi} \sim \epsilon^{-1/2}$;  hence the timescale $\tau \sim \epsilon^{-1/2} \omega_{\rm tide}^{-1}$ and Eq.~\eqref{mathis_ratio} can be recast as 
\begin{equation}
    \frac{\nu_{\rm turb \, v}}{\nu_{\rm turb \, h}} \sim  \frac{2 \, \epsilon \,\Omega^{2} \,  \omega_{\rm tide}^{2}}{N^{4}}. 
    \label{mathis_ratio1}
\end{equation}
To illustrate the application of this formula, we consider two different limiting cases. 
First, we suppose that the downdrafts can penetrate undisturbed below the convection zone and enforce a nearly adiabatic stratification as given by Eq.~\eqref{subadiabaticity}. Assuming, for example, $\nabla_{\rm ad}-\nabla \sim 10^{-5}$ in the layer and adopting the characteristic parameters at the base of the convection zone of WASP-18 (see Sect.~\ref{stellar_model}), that is, $g=320$~m~s$^{-2}$, and $H_{\rm p}=4.4 \times 10^{7}$~m, we find $N = 8.5 \times 10^{-6}$~s$^{-1}$ and $\nu_{\rm turb\, v}/\nu_{\rm turb\, h} \sim 0.1$ from Eq.~\eqref{mathis_ratio1} where $\epsilon =\epsilon_{\rm b} \sim 10^{-4}$ for our system (cf. Sect.~\ref{interaction}). Such a value  indicates that the vertical turbulent diffusivity is comparable with the horizontal diffusivity as a consequence of the nearly adiabatic stratification of the overshoot layer where  $N$ is very small. This conclusion does not depend strongly on the adopted value of $\nabla_{\rm ad} - \nabla$, that is uncertain, but on the fact that the stratification is close to adiabatic. 

Secondly, we assume that the  turbulence produced by the tidal flow is capable of disrupting the downdrafts so that they cannot modify the stratification in the top of the radiative zone. Therefore,  $\nabla_{\rm ad} - \nabla \sim 0.04$ and $N = 5.6 \times 10^{-4}$~s$^{-1}$ at a depth of $0.1\, H_{\rm p}$ below the base of the stellar convection zone of our model of WASP-18 (cf. Sect.~\ref{stellar_model}, Fig.~\ref{fig1}) giving $\nu_{\rm turb\, v}/\nu_{\rm turb\, h} \sim 6 \times  10^{-9}$ from Eq.~\eqref{mathis_ratio1}. In this regime, the turbulent diffusivity is essentially horizontal because the vertical turbulent motions are strongly hindered by the stable sub-adiabatic stratification and do not contribute in any appreciable way to the diffusivity. 

In conclusion, it seems better to adopt a conservative approach and consider only the horizontal diffusivity to estimate the diffusion timescale of the downdrafts $\tau_{\rm d}$. This choice provides an upper limit for its value in the first case, while giving its order of magnitude in the second case. Specifically, we write the downdraft diffusion timescale as
\begin{equation}
    \tau_{\rm d} \sim \frac{b^{2}}{\nu_{\rm turb \, h}},
\label{time_dect}    
\end{equation}
where $b$ is the horizontal size of the downdrafts at the top of the overshoot layer as given by Eq.~\eqref{downdraft_rad}, while the horizontal turbulent kinematic diffusivity is given by Eq.~\eqref{hor_turb_1}. 

\subsection{Modifying the stratification in the overshoot layer due to tidally induced turbulence}
\label{modification}

We assume that the downdrafts are unable to significantly affect the stratification of the radiative layers below the convection zone whenever $\tau_{\rm d} \ll \tau_{\rm p}$, where $\tau_{\rm p}$ is the characteristic lifetime of the downdrafts themselves in the absence of tides as given by Eq.~\eqref{tau_downdraft}. In other words, in this regime, the turbulent diffusivity induced by the tides strongly limits the lifetime of the downdrafts and disrupts the process leading to a nearly adiabatic stratification in the overshoot layer \citep{Zahn91}. 
On the other hand, when $\tau_{\rm d} \gg \tau_{\rm p}$, we assume that the downdrafts are unperturbed by the equilibrium tidal flow so they can enforce a nearly adiabatic stratification in the overshoot layer as predicted by the model by \citet{Zahn91}.

In the intermediate regime when $\tau_{\rm d}$ is comparable with $\tau_{\rm p}$, we assume that the number of downdrafts ${\cal N}$ and their filling factor $f \propto {\cal N}$ are affected. This in turn changes the sub-adiabatic gradient because $\nabla_{\rm ad} - \nabla$ is proportional to $f$ according to Eq.~\eqref{subadiabaticity}. 
A quantification of the variation in the filling factor can be derived by considering that  $\cal N$  at any given time obeys the equation 
\begin{equation}
\frac{d {\cal N}}{dt} = \frac{d{\cal N}_{\rm c}}{dt} + \frac{d{\cal N}_{\rm d}}{dt},
\label{downdraft_number}
\end{equation}
where $d{\cal N}_{\rm c}/dt$ is the rate of formation of new downdrafts  per unit time, that we assume to be a constant $k_{\rm c}$, while $d{\cal N}_{\rm d}/dt$ is the rate of destruction of the downdrafts due to their intrinsic decay and the effect of the turbulent diffusion as parameterised by Eq.~\eqref{time_dect}. 

We may write $d {\cal N}_{\rm d}/dt = - {\cal N}/\tau_{0}$, where $\tau_{0}$ is a characteristic decay time that  in the absence of any tidally induced turbulence is $\tau_{0} = \tau_{\rm p}$, where $\tau_{\rm p}$ is the intrinsic downdraft lifetime as given by Eq.~\eqref{tau_downdraft}. On the other hand, considering  a turbulent diffusion process with the timescale $\tau_{\rm d}$ as given by Eq.~\eqref{time_dect} that adds its effect on the intrinsic decay of the downdraft, we find a decay of the vertical flow of the downdraft proportional to $\exp(-t/\tau_{0})$, where 
\begin{equation}
\frac{1}{\tau_{0}} = \frac{1}{\tau_{\rm p}} + \frac{1}{\tau_{\rm d}}, 
\label{total_diff_time}
\end{equation}
as we show in Appendix~\ref{appendixA}. 
The stationary solution of Eq.~\eqref{downdraft_number} gives the  number of downdrafts that straddle the overshoot layer as ${\cal N} = k_{\rm c} \tau_{0}$, hence we can assume that $f \propto k_{\rm c} \tau_{0}$. 
Having quantified the dependence of $f$ on the decay timescale $\tau_{0}$ of the downdrafts (see Sect.~\ref{application}) we  were then able to compute the impact of the additional decay produced by the tidally induced turbulence on the sub-adiabatic gradient in the overshoot layer of WASP-18 via Eq.~\eqref{subadiabaticity}. 

{Our analysis  has been restricted to the portion of a downdraft straddling the overshoot layer assuming that the part of the same downdraft located in the overlying convection zone is not significantly perturbed by the equilibrium tidal flow. We go on to provide an argument in support of this assumption. %This is actually the case because the turbulent diffusivity in the convection zone  is much higher than that produced by the mechanism we propose owing to the convective instability of that zone of the star (cf. Sect.~\ref{stellar_model}). 

The oscillating equilibrium tidal flow has a period much shorter than the typical convective turnover time in the bulk of the convection zone and close to its base. Therefore, we are in the regime where the  model proposed by \citet{Terquem23} can be applied to estimate the  kinetic energy  per unit mass and time exchanged  between the tidal flow and an average convective downdraft. From that estimate, we derive a characteristic timescale for the kinetic energy exchange between the tidal flow and a downdraft of $\sim 2 \times 10^{7}$~s in the case of WASP-18, that is, one order of magnitude longer than the  mean lifetime of the downdrafts as evaluated in Sect.~\ref{stellar_model}. Therefore, we do not expect a significant influence of the tidal flow on the dynamics of the convective downdrafts in the bulk of the convection zone. The same conclusion is true for the influence of the tidal flow on a stellar dynamo working in the bulk of the convection zone because of the very small contribution of the tidal flow to the shear implying that it cannot appreciably affect the amplification, the turbulent diffusion, and the magnetic buoyancy of the magnetic field there (cf. Sect.~\ref{dynamo_models} and the final paragraph of Appendix~\ref{app_continuous_layer}).  }

\subsection{Consequence for the storage of magnetic flux in the overshoot layer}
\label{magn_overshoot}

We adopt the paradigm proposed by \citet{Moreno-Insertisetal92}, \citet{Ferriz-MasSchussler94}, \citet{Ferriz-Masetal94}, \citet{Caligarietal95,Caligarietal98}, where the magnetic flux in the overshoot layer is organised into slender magnetic flux tubes that are initially stored inside the layer. Their field is steadily amplified by the radial shear of  the tachocline until its intensity reaches the threshold for the onset of an undulatory instability that leads the flux tubes to emerge into the convection zone and reach the photosphere where they produce active regions. 

The formation of slender magnetic flux tubes from an initial nearly uniform layer of magnetic azimuthal flux can be a consequence of the penetrative convection and of magnetic field instabilities as discussed, for instance, by  \citet{Fan21}. 
Although some difficulties have been put forward for the operation of such a paradigm in the solar and stellar convection zones and overshoot layers \citep[cf. Sect.~11 in][]{Fan21,Hughes07}, we consider such a model in view of its simplicity, in particular the possibility of analytically predicting the instability of the slender magnetic flux tubes in the overshoot layer thanks to the approach introduced by \citet{Ferriz-MasSchussler94}. It is interesting to note that the storage and emergence of flux tubes from the overshoot layer of the Sun have recently been modelled by \citet{Maneketal22} to account for the observed hemispheric dependence of the sign of magnetic helicity in solar active regions, thus providing further support to our assumptions about the key role of the overshoot layer in stellar activity. 

When the downdrafts are allowed to establish a nearly adiabatic stratification in the overshoot layer, the sub-adiabatic gradient is such that $\nabla_{\rm ad} -\nabla \sim (0.1-1) \times 10^{-5}$. In this regime, the slender magnetic flux tubes stored and amplified within the layer become unstable with a growth timescale of the order of one year for a field intensity of $\sim 10$~T \citep[cf.][]{Ferriz-MasSchussler94} and, after penetrating in the overlying convection zone, emerge almost radially forming active regions at low latitudes in the stellar photosphere \citep{Caligarietal95,Granzeretal00}. 
On the other hand, when the convective downdrafts are made ineffective by the turbulent diffusivity due to tides, the stratification in the former overshoot layer becomes  that of the radiative interior with a much larger sub-adiabatic gradient difference $\nabla_{\rm ad} -\nabla \sim 0.04$ (see Sect.~\ref{stellar_model}, Fig.~\ref{fig1}). This implies that a field intensity much larger  than $10$~T is required to produce the flux tube instability, that {we assume to be} impossible to attain for the stellar dynamo because it is quenched by the magnetic field itself \citep[cf.][]{MorenoInsertisetal95,GilmanRempel05}. As a consequence, the slender flux tubes remain stably stored inside the overshoot layer and cannot emerge to form active regions in the photosphere.

In the intermediate regime when $\tau_{\rm p} \sim \tau_{\rm d}$, we have a decrease in the filling factor, $f,$ of the downdrafts because we know that $f \propto k_{\rm c} \tau_{0}$ (Sect.~\ref{modification}). As a consequence, the gradient becomes more sub-adiabatic, that is, the difference $\nabla_{\rm ad} - \nabla$ increases because it is inversely proportional to $f$ (cf. Eq.~\ref{subadiabaticity}). Therefore,  the undulatory instability of the slender flux tubes stored in the overshoot layer is  hampered because it requires a remarkably stronger field to develop. For example, looking at Fig.~8 of \citet{Ferriz-MasSchussler94}, we see that the threshold for the development of the instability is at a field intensity of $13.5$~T when $\nabla_{\rm ad}- \nabla = 10^{-6}$ in the Sun, while it increases to $15 $~T when $\nabla_{\rm ad}- \nabla = 2 \times 10^{-6}$ that corresponds to a decrease in $f$ by a factor of 2 that is obtained when $\tau_{\rm p} \sim \tau_{\rm d}$.
Such a stronger magnetic field intensity may not be reachable through the dynamo amplification leading to the same situation described above in the case of the downdraft disruption. In other words, the magnetic flux tubes may not be capable of developing the instability that allows them to emerge to the photosphere. This will strongly reduce the level of stellar activity. As we show in the next section (Sect.~\ref{application}), this could be  the case in WASP-18, thus providing a possible explanation for its depressed activity level.

{The above conclusion is based on the assumed filamentary structure of  the toroidal field in the overshoot layer. If the field is structured as  a continuous layer rather than as  discrete individual slender flux tubes, it is important to take into account other kinds of instabilities that are specific to such a layer. Previously, we pointed out that the magnetic pumping by the downdrafts should lead to the formation of such a layer and confine it  to the lower part of the overshoot region. Therefore, we present an analysis of the relevant instabilities in a continuous magnetic layer in Appendix~\ref{app_continuous_layer}. It confirms the conclusion we reached by means of the slender flux tube model, that is, also a continuous magnetic layer is stabilised by an increase in the sub-adiabatic gradient in the overshoot region below the convection zone. }

   \begin{figure}
%   \centering
\hspace{-1.7cm}
   \includegraphics[width=10cm,height=12cm,angle=270]{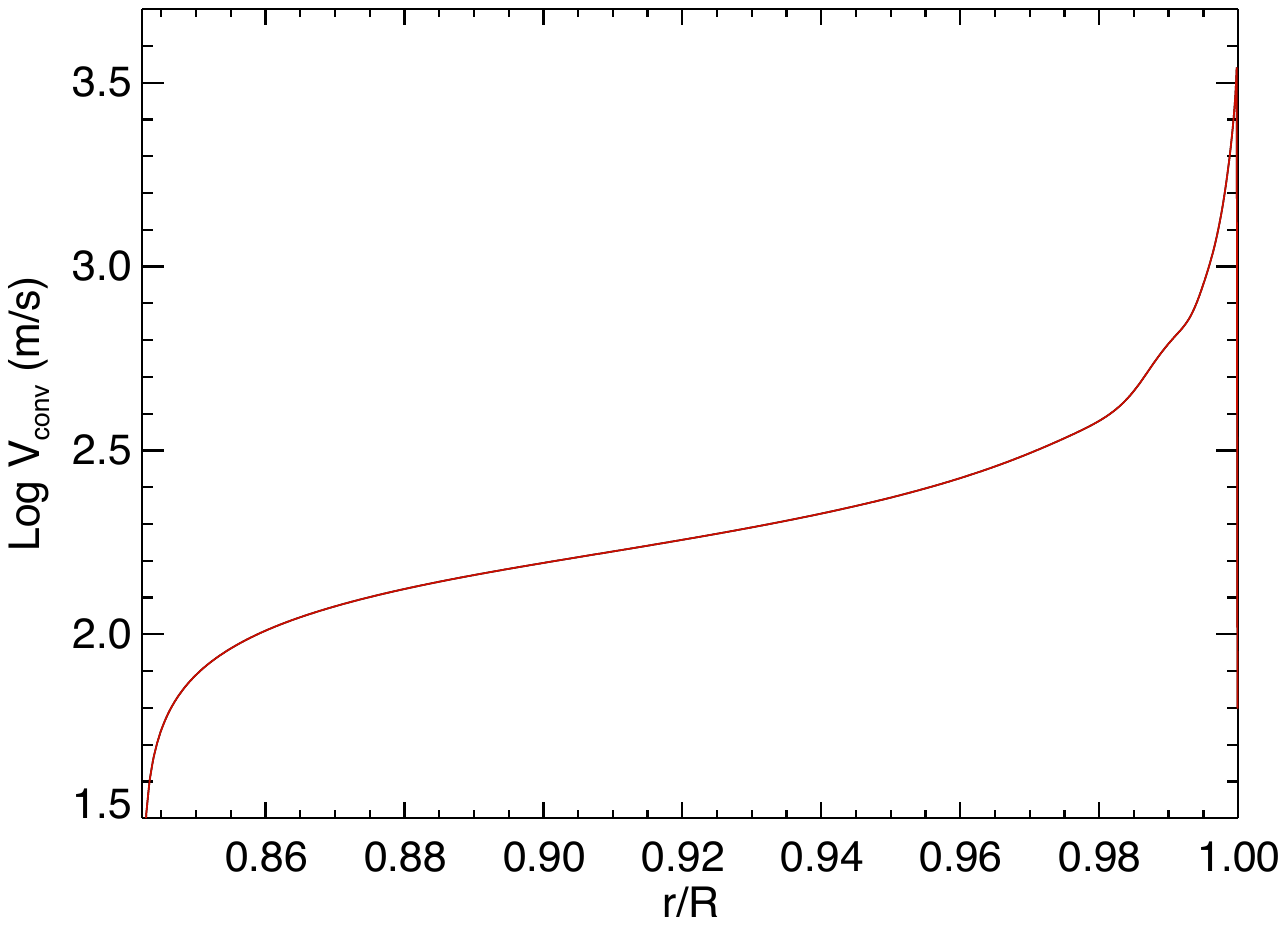}
   \vspace{-1.0cm}
   \caption{Logarithm of the convective velocity $V_{\rm conv}$ vs. the stellar radius $r$ in units of the photospheric radius $R$ in the outer convection zone of our interior model (red solid line). }
   \label{fig_cv}
    \end{figure}

\section{Application to the WASP-18 system}
\label{application}
In order to make a quantitative application of the model proposed above to the WASP-18 system, we need a model of the star internal structure that is described in the next  Sect.~\ref{stellar_model}. Using that model, we compute estimates for the turbulent diffusivity and study the regime occurring in the stellar overshoot layer and its possible consequences on stellar activity in Sect.~\ref{model_application}. 

\subsection{Stellar interior model}
\label{stellar_model}
A stellar interior and evolution model has been computed by means of the Modules for Experiments in Stellar Astrophysics (MESA) code \citep[][and references therein]{Paxtonetal19} that has been run through the web interface MESA-Web\footnote{http://user.astro.wisc.edu/~townsend/static.php?ref=mesa-web}. The model has been computed for a non-rotating star of initial mass of 1.22~M$_{\odot}$ and a solar metallicity ($Z=0.02$) with the basic network for nuclear reactions; the mixing-length parameter $\alpha_{\rm mlt} = 2.0$, while the overshooting parameters were assumed to be very small \citep[$  f=10^{-4}$ and $f_{0}=5 \times 10^{-4}$; these parameters control the length scale of the exponential decay of the diffusive overshooting and the lower boundary of the overshoot region, respectively, and are not to be confused with the downdraft filling factor $f$ considered in the present work; see][Sect.~3, for their definition]{Moravvejietal16}. We note that such overshooting parameters have an impact on the internal diffusion mainly during late evolutionary phases of the stellar model \citep[cf. Sect.~5.2 of][]{Paxtonetal11}, so they do not affect our application for which we may assume an internal structure model computed without a significant overshooting at the base of the outer convection zone. We selected the interior structure model corresponding to an age of 662.5~Myr, when the star has a radius $R= 1.208$~R$_{\odot}$ that matches closely the  radius of WASP-18 \citep[$R=1.216 \pm 0.067$~R$_{\odot}$;][]{Hellieretal09}. 
The base of the convection zone  determined according to the Schwarzschild criterion is at $r_{\rm b}/R= 0.8438$ and the convective velocity in that layer is 45.9~m~s$^{-1}$; the pressure scale height there is $H_{\rm p} = 4.376 \times 10^{7}$~m, the density  $\rho_{\rm b} = 10.4$~kg~m$^{-3}$, and the temperature  1.054~MK. 

In Fig.~\ref{fig_cv}, we plot the convective velocity $V_{\rm conv}$ as computed from the mixing-length theory vs. the radius in the convection zone of our interior model. The rapid decrease in the density towards the photosphere makes the convective velocity increase remarkably in order to transport the whole stellar luminosity that amounts to 2.0~L$_{\odot}$ in our model. The convective turnover time at the base of the convection zone, that is adopted as the downdraft lifetime in our model, is  $\tau_{\rm p} \sim 1.8 \times 10^{6}$~s or $\sim 21$~days by applying Eq.~\eqref{tau_downdraft}. On the other hand, the asteroseismic calibration by \citet{Corsaroetal21}, indicates an average convective turnover time of $\sim 29 \pm 8$~days for a MS star with $B-V=0.49$, the colour index of WASP-18. Therefore, our estimate of the convective turnover time at the base of the convection zone, and consequently of $\tau_{\rm p}$,  should likely be regarded as a lower limit for our star. 

\begin{figure}
%   \centering
\hspace{-1.7cm}
   \includegraphics[width=10cm,height=12cm,angle=270]{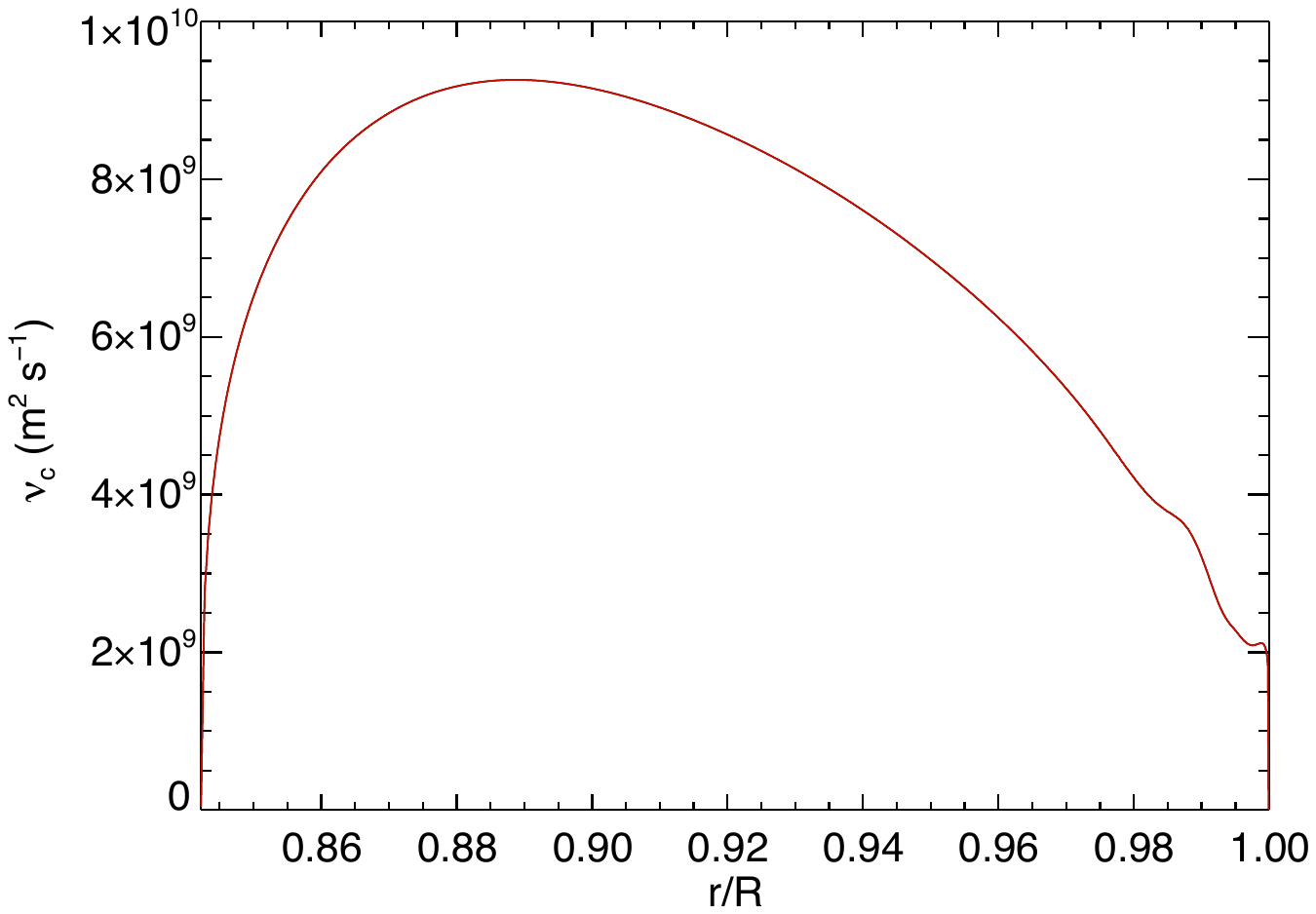}
   \vspace{-1.0cm}
   \caption{Turbulent convective diffusivity $\nu_{\rm c}$ vs. the stellar radius $r$ in units of the photospheric radius $R$ as derived from our stellar interior model (red solid line).}
   \label{conv_diffus}
    \end{figure}

In Fig.~\ref{conv_diffus}, we plot the convective turbulent diffusivity $\nu_{\rm c} \sim \alpha_{\rm mlt} H_{\rm p}(r) V_{\rm conv}(r)$ vs. the stellar radius $r$. It is useful for a comparison with the diffusivity produced by the turbulence induced by the tidal flow according to our mechanism (see Sect.~\ref{model_application}). The value of $\nu_{\rm c}$ averaged over the whole volume of the convection zone is  $\overline{\nu}_{\rm c} \sim 7.0 \times 10^{9}$~m$^{2}$~s$^{-1}$. 
%%%%%%%%%%%%%%%%%%%%%%%%%%%%%
%%%%%%%%%%%%%%%%%%%%%%%%%%%%%%

\begin{figure}
%   \centering
\hspace{-1.0cm}
   \includegraphics[width=10cm,height=12cm,angle=270]{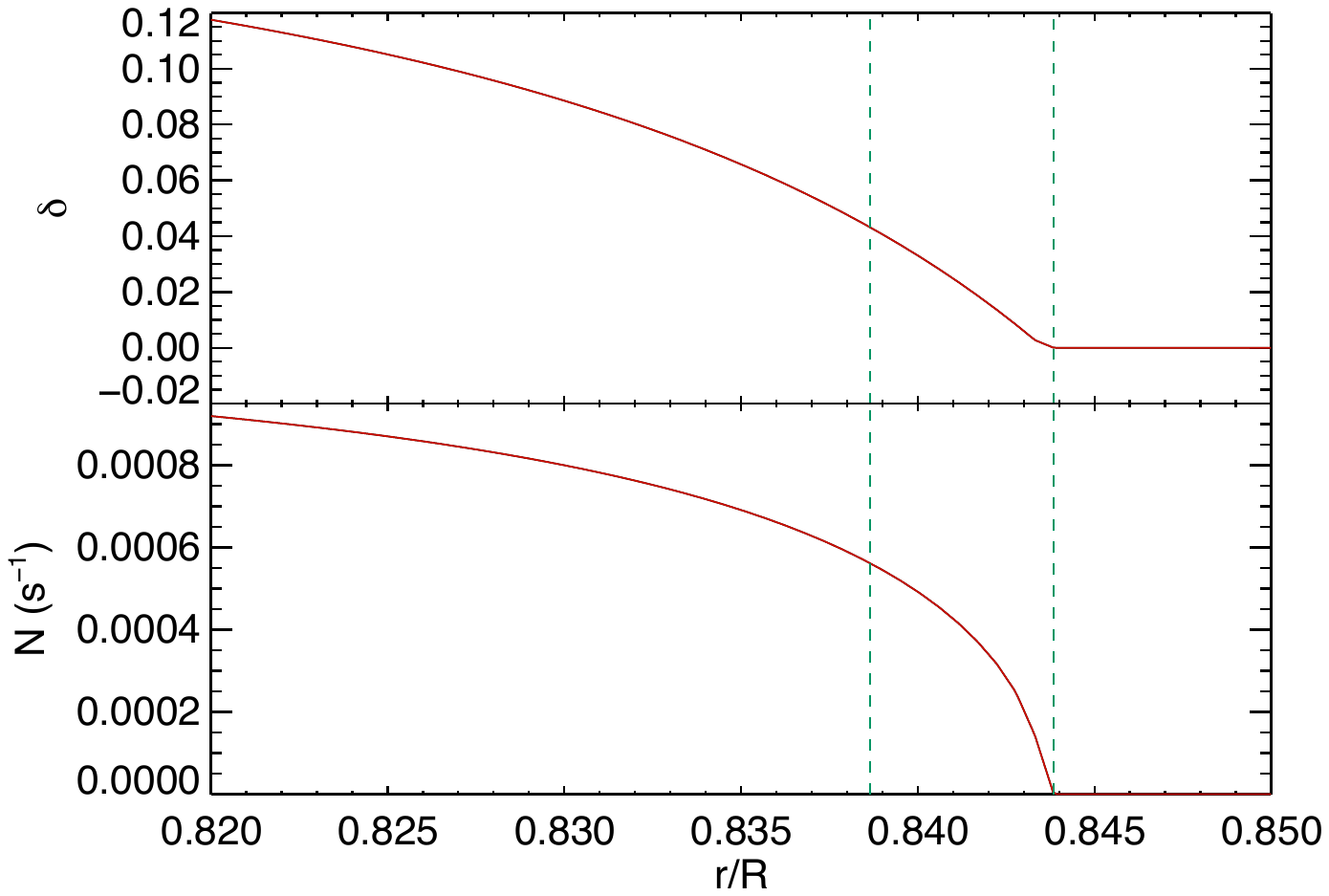}
   \vspace{-1.0cm}
   \caption{Difference of $\delta = \nabla_{\rm ad} - \nabla$ between the adiabatic thermal gradient and the actual stellar thermal gradient (upper panel) and the Brunt-V\"ais\"ala frequency $N$ (lower panel) vs. the stellar radius $r$ in units of the photospheric radius $R$ (red solid lines). The vertical green dashed lines bound a layer of thickness $L_{\rm p} = 0.1\, H_{\rm p}$, where $H_{\rm p}$ is the pressure scale height at the base of the stellar convection zone.}
   \label{fig1}
    \end{figure}

In Fig.~\ref{fig1}, we plot the difference between the adiabatic and the actual stellar thermal gradient $\delta \equiv \nabla_{\rm ad} - \nabla$ vs. the stellar radius close to and below the base of our model convection zone. The base of the stellar convection zone according to the Schwarzschild criterion is indicated by the right vertical green dashed line where $\delta =0$, while the left green vertical dashed line is drawn below a depth of $0.1\, H_{\rm p}$ from the base of the convection zone, where $H_{\rm p}$ is the pressure scale height at the base of the convection zone. Such a depth is assumed to be the vertical extension of the overshoot layer,  $L_{\rm p}$, that is, the penetration depth of the convective downdrafts. The difference between the adiabatic gradient and the actual temperature gradient increases rapidly  as one moves into the radiative interior because our model has been computed without a significant amount of overshooting, thus it does not include the effects of the downdrafts on the thermal stratification. 
%%%%%%%%%%%%%%%%%%%%%%%%%%%%

\begin{figure}
%   \centering
\hspace{-1.7cm}
   \includegraphics[width=10cm,height=12cm,angle=270]{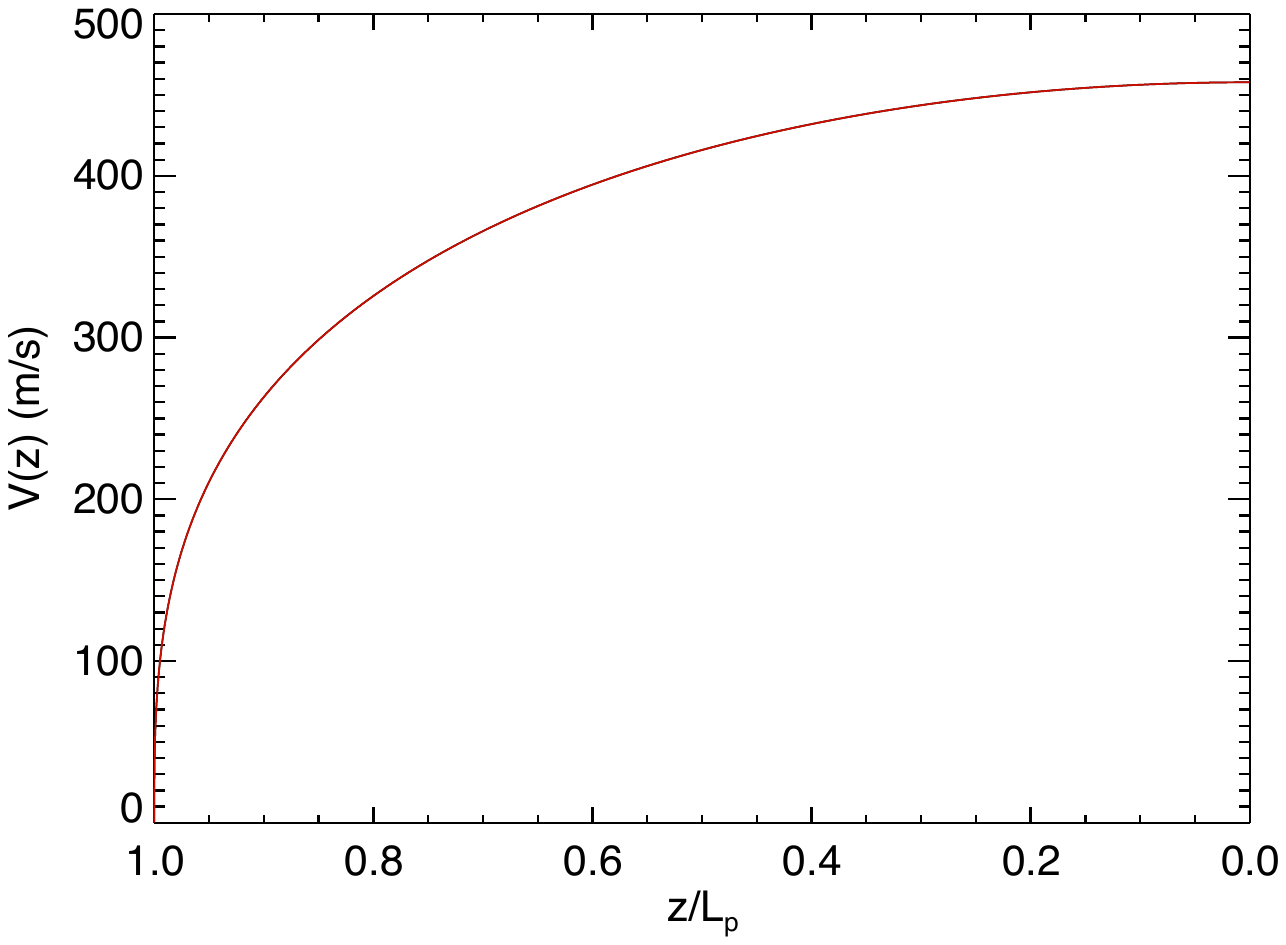}
   \vspace{-1.0cm}
   \caption{Velocity of the downdrafts in the overshoot layer of our model vs. the depth $z$  below the base of the convection zone measured in units of $L_{\rm p}$, that is, the penetration of the plumes below the standard convection zone. }
   \label{fig_downdraft}
    \end{figure}

The depth, $z_{0}$, of  the convection zone in our model of WASP-18 is 0.156~stellar radii or $1.31 \times 10^{8}$~m. The corresponding radius of the convective downdrafts at the base of the convection zone, computed by means of Eq.~\eqref{downdraft_rad}, is $b= 6.6 \times 10^{6}$~m, while the initial velocity of penetration $V_{\rm b} = 453$~m~s$^{-1}$ according to Eq.~\eqref{downdraft_vel}. A plot of the penetration velocity vs. the depth expressed as a fraction of the penetration depth, $L_{\rm p}$, is shown in Fig.~\ref{fig_downdraft} and has been computed by means of Eq.~\eqref{downdraft_vel1}. We note the fast braking of the plumes as they approach the penetration depth $L_{\rm p}$ where there is a thin boundary layer separating the overshoot layer from the radiative zone below \citep[see Sect.~3.2 of][for details]{Zahn91}. 

Our interior model together with the above estimate of the initial velocity of the downdrafts can be used to evaluate the deviation of the stratification from the adiabatic gradient in the overshoot layer by means of Eq.~\eqref{subadiabaticity}, giving $\nabla_{\rm ad} - \nabla \sim 6.5 \times 10^{-6}$ for a plume filling factor $f=0.1$. This result shows that the downdrafts are very efficient in bringing the stratification very close to a perfectly adiabatic one. The gradient, however, remains slightly sub-adiabatic, that is crucial  to allow the storage of magnetic flux tubes and set the threshold for their instability  \citep[cf.][]{Ferriz-MasSchussler94}. 

\subsection{Turbulence in the overshoot layer and its effects on the downdrafts and stratification}
\label{model_application}

As we show in Sect.~\ref{interaction}, the close-by and massive planet WASP-18b produces a tidal deformation of $\epsilon_{\rm b} \sim 1.0\times 10^{-4}$ and a horizontal tidal velocity of $\sim 18$~m~s$^{-1}$ in the overshoot layer. 
Applying Eq.~\eqref{hor_turb_1} with $\omega_{\rm tide} = 1.3 \times 10^{-4}$~s$^{-1}$, we find a horizontal turbulent diffusivity in that layer of $\nu_{\rm turb \, h} \sim 4 \times 10^{8}$~m$^{2}$~s$^{-1}$. The  diffusion timescale of a downdraft of size $b=6.6 \times 10^{6}$~m is of  $\tau_{\rm d} \sim 1.1  \times 10^{5}$~s according to Eq.~\eqref{tau_downdraft}, that is, a factor of $\sim 16$  shorter than the downdraft lifetime $\tau_{\rm p}$ as estimated from our interior model and Eq.~\eqref{time_dect}. Adopting the average convective turnover time estimated with the calibration by \citet{Corsaroetal21} as the downdraft lifetime, we obtain a ratio  $ \tau_{\rm p}/\tau_{\rm d} 
 \sim 23 \pm 7$. 

 According to the simple model introduced in Sect.~\ref{modification}, we have a value of $\tau_{0}$ that is reduced  by a factor of $\sim 25$ with respect to the case without equilibrium tide. The corresponding filling factor $f$ of the downdrafts that appears in Eq.~\eqref{subadiabaticity} is therefore reduced by the same factor that implies an increase in the sub-adiabaticity $\nabla_{\rm ad} - \nabla$ by a factor of $\sim 25$, that is, from $\nabla_{\rm ad} -\nabla \sim 6.5 \times 10^{-6}$ (cf. Sect.~\ref{stellar_model}) to $\nabla_{\rm ad} - \nabla \sim 1.6 \times 10^{-4}$. As a consequence, the slender magnetic flux tubes stored in the overshoot layer must reach  a remarkably larger field strength to become unstable and emerge to the stellar photosphere. 
 
\begin{figure}
%   \centering
\hspace{-0.75cm}
\vspace{0.0cm}
   \includegraphics[width=8cm,height=11cm,angle=270]{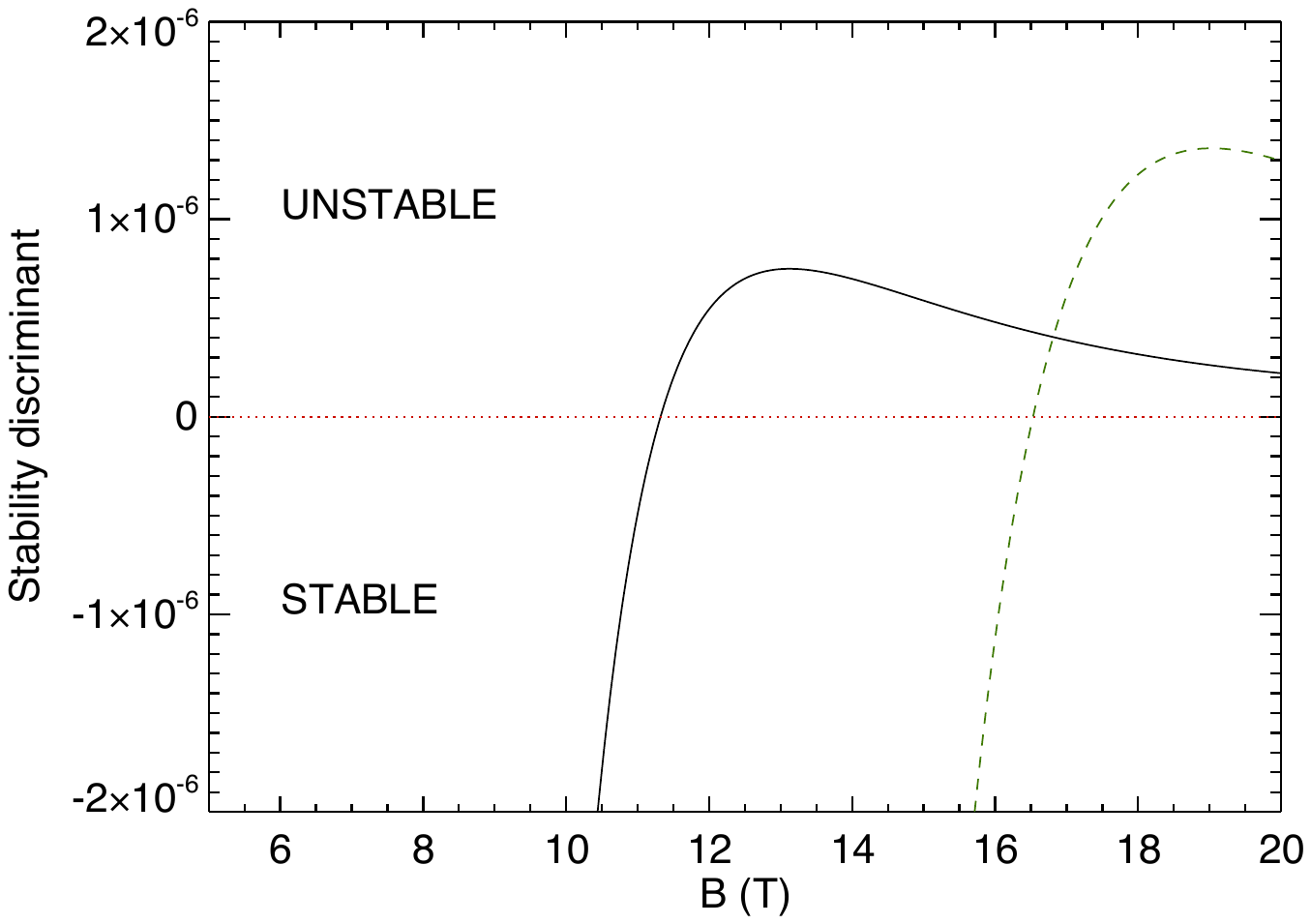}
   \vspace{-0.5cm}
   \caption{Stability diagram of a slender flux tube to the $m=1$ mode computed according to  \citet{Ferriz-MasSchussler94} for two different values of the sub-adiabatic gradient in the overshoot layer $\nabla_{\rm ad} - \nabla=$ $6.5 \times 10^{-6}$ (black solid line) and $1.6 \times 10^{-4}$ (green dashed line). The horizontal red dotted line marks the transition from a stable to an unstable flux tube (see  text). }
   \label{fig_stability}
    \end{figure}

 In Fig.~\ref{fig_stability}, we show the stability diagram of such  flux tubes  computed by means of Eq.~(40) of \citet{Ferriz-MasSchussler94} for the most unstable mode with azimuthal wavenumber $m=1$. The radial differential rotation parameter $q=0.06$, that is, similar to the solar value, while the difference between the angular velocity of the plasma outside and inside the flux tube is assumed to be maximal, that is, equal to $\varv_{\rm A}/r_{\rm b}$, where $\varv_{\rm A}$ is the Alfven velocity inside the flux tube \citep[see][for the definition and role of these parameters]{Ferriz-MasSchussler94}. We consider two values of the sub-adiabatic gradient, that is, $\nabla_{\rm ad} - \nabla = 6.5 \times 10^{-6}$, that corresponds to the stratification in the overshoot layer of WASP-18 without the perturbation due to the tides, and $\nabla_{\rm ad} - \nabla = 1.6 \times 10^{-4}$, that corresponds to the increased sub-adiabaticity produced by the reduced efficiency of the downdrafts under the action of the equilibrium tide. As can be seen from Fig.~\ref{fig_stability}, the threshold field for the instability  of the flux tubes in the overshoot layer of WASP-18 increases from $\sim 11.5$~T to $16.5$~T when we consider the effect of the tidal flow on the downdrafts. This corresponds to an increase in the magnetic energy density by a factor of $\sim 2$. If the stellar hydromagnetic dynamo is unable to produce a field strength that exceeds such an increased instability threshold, the magnetic flux tubes stored in the stellar overshoot layer cannot emerge and the magnetic activity of WASP-18 is depressed providing a possible explanation for the observations. 

%In such a regime, we assume that the downdrafts  are ineffective in extending the adiabatic stratification below the convection zone, so that the thermal gradient becomes radiative immediately below the base of the convection zone as illustrated in Fig.~\ref{fig1}. Considering the stratification at a depth of $0.1 \, H_{\rm p}$ below the convection zone, that is, the adopted  lower boundary of the overshoot layer in the absence of tidal effects, we see that $\nabla_{\rm ad} - \nabla \sim 0.04$ there. 
%Such a rather high sub-adiabaticity would require a magnetic field strength much larger than  $\sim 10^{5}$~G  to produce the instability and the emergence of the magnetic slender flux tubes that are responsible for photospheric activity in the model we discussed in Sect.~\ref{magn_overshoot}. 

We could consider a slow inflow of heat due to  radiative conductivity, bringing a flux tube into thermal equilibrium with its surroundings, as a possible way of reducing flux tube stability because a flux tube in thermal equilibrium would become buoyantly unstable. However, the timescale for such a mechanism to operate is of the order of $L_{\rm p}^{2}/K$, assuming the flux tubes have a diameter comparable with the vertical extension of the overshoot layer. 
From our model we find, $L_{\rm p} \sim 4.4 \times 10^{6}$~m and $K \sim 3.9 \times 10^{4}$~m$^{2}$~s$^{-1}$ that gives a thermal diffusion time of the order of $\sim 15$~years, that is, remarkably longer than the typical short stellar activity cycles observed in rapidly rotating F-type stars having a length of $\sim 0.3-3$ years \citep[cf.][]{Mittagetal19}. Therefore, we rule out such a mechanism as a source of flux tube destabilisation because it cannot operate over timescales shorter than the stellar cycles as required to account for the activity of such stars. 

{Finally,  in Appendix~\ref{app_continuous_layer}, we consider the case when the magnetic flux in the overshoot region is organised as a continuous magnetic layer. We find that an increase in the sub-adiabatic stratification leads to a remarkably larger threshold for the field instability also in that case, notably also considering  the doubly diffusive instability or the overstability that are not included in the model by \citet{Ferriz-MasSchussler94}. Therefore, the above conjecture  about the suppression of magnetic activity in WASP-18 by the action of the equilibrium tide is not critically dependent on our assumption on the field organisation in the overshoot layer. }% where the lifetimes of active regions are on the order of weeks or months at most.  
%%%%%%%%%%%%%%%%%%%%%%%%%%%%%%%%%%%%%%
%%%%%%%%%%%%%%%%%%%%%%%%%%
%We conclude that the surface activity of WASP-18 can be driven only by a dynamo  working in the bulk of its convection zone without the contribution of flux tubes stored inside the overshoot layer. The timescale for the turbulent diffusion of starspots in the photosphere of WASP-18 under the action of the equilibrium tidal flow can be estimated from the considerations in Sect.~\ref{spot_diffusion} giving $\tau_{\rm spot} \sim 10^{5}$~s for $r_{\rm spot} \sim 10^{7}$~m. Such a timescale is shorter than or comparable with the typical timescales of sunspot formation, suggesting that the process of spot formation can be severely  hampered in the photosphere of WASP-18. Therefore, we expect that only small-scale and rather diffuse magnetic fields can be present in the photosphere of such a planet host accounting for its very low activity level.  

\subsection{The case of a more distant or lighter planet}

WASP-18b was probably formed at a larger distance than the present separation where it was likely brought by the tides it raised on the star \citep[e.g.][]{CollierCameronJardine18}. Considering a past separation of twice the present value, the deformation parameter $\epsilon$ becomes 8 times smaller and the tidal frequency  about 2.8 times smaller leading to an horizontal tidal velocity of only 0.8~m~s$^{-1}$ and a turbulent horizontal diffusivity as estimated from Eq.~\eqref{hor_turb_1} of $\nu_{\rm turb \, h} \sim 4 \times 10^{6}$~m$^{2}$~s$^{-1}$ (cf. the dependence of $\nu_{\rm turb \,h}$ on $(\omega_{\rm tide} \epsilon_{\rm b}) ^{3/2}$ in Eq.~\ref{hor_turb_1}). The diffusion timescale of the downdrafts becomes  $\tau_{\rm d} \sim 1.1 \times 10^{7}$~s, that is, about a factor of $\sim 4-5$ times longer than the typical downdraft lifetime. Therefore, a modest modification is expect to occur in the overshoot layer stratification as a consequence of the equilibrium tide raised by the planet.  We conclude that WASP-18 has been caught in a particular phase of its evolution when the proximity and the large mass of its planet can modify the stratification in its overshoot layer, possibly hampering its magnetic activity.

Similar conclusions apply if we consider a planet of lower mass on the same close-by orbit. For example, in the case of \object{WASP-12}, the hot Jupiter has a mass of $1.36$ Jupiter masses and an orbital separation of $0.023$~au comparable with that of WASP-18b. In that system, we expect an increase in the diffusion time by a factor of $\sim 30$ with respect to WASP-18 because of the dependence on $\epsilon_{\rm b}^{3/2}$ of the turbulent diffusivity. Therefore, the decrease in $\tau_{0}$ and consequently in $f$ is only by a factor of $\sim 2$, that should not produce a significant effect on the flux tube stability and stellar activity. In conclusion, the apparent depressed chromospheric activity of WASP-12 should be interpreted in the framework of a  circumstellar absorption model, as mentioned in Sect.~\ref{introduction}.

\section{Discussion and conclusions}
\label{conclusions}

We propose a {conjectural model} to account for the very low level of magnetic activity observed in WASP-18 based on the disruption of the convective plumes in its overshoot layer and the consequent hindering of the process of starspot formation in its photosphere. Such a disruption is a consequence of the interaction of the hori\-zon\-tal tidal flow with the vertical obstacles represented by the convective downdrafts in the overshoot layer that leads to an enhanced turbulent diffusion  with respect to the case of a star without tidal interactions. 

{We have presented plausibility arguments to support the hypotheses upon which our model is based. Admittedly, most of them rely on simplified analytical models of physical processes that are very difficult or impossible to simulate numerically in the regimes characteristic of stellar interiors. Therefore, our proposal must be regarded as speculative and is presented in order to stimulate further observational and theoretical investigations on the effects of close-by massive planets on stellar magnetic activity.}

The mechanism proposed in this paper can work only in a restricted range of mass and separation of the close-by planet, while a basic requirement for its operation is a large tidal frequency, $\omega_{\rm tide}$, which is possible only if the stellar host is not synchronised. Therefore, our mechanism cannot operate in close stellar binary systems because both their components are synchronised. In the case of WASP-18, our model suggests that the inhibition of stellar activity has begun only rather recently in the evolution of the system because the invoked effect was not relevant when the separation of the planet was twice the present one. The timescale for the orbital decay of the system depends on the ill-known modified tidal quality factor of the star that can be as large as $Q^{\prime}_{\rm star} \ga 2 \times 10^{8}$ in a system away from synchronisation as WASP-18 leading to a remaining lifetime before engulfment of the order of 100~Myr, despite the very close orbit of the planet \citep[cf.][]{OgilvieLin07,Mathis15,Bonomoetal17,CollierCameronJardine18}. 

The additional turbulent dissipation invoked in our model has a negligible effect on the dissipation of the equilibrium (and possibly dynamical) tide because it is  smaller by a factor of $\sim 15-20$ than the average turbulent diffusivity $\overline{\nu}_{\rm c}$ due to the convective motions in the stellar convection zone (cf. Sect.~\ref{stellar_model}) that are regarded as the main source of dissipation of the equilibrium tide \citep{Zahn77}\footnote{For simplicity, we neglect the reduction of the convective turbulent viscosity acting on the equilibrium tide that occurs when the tidal frequency is significantly larger than the inverse of the convective turnover time as given by the mixing-length theory in a given layer \citep[e.g.][Sect.~2.1.1]{Barker20}.}. Moreover, it is limited to the shallow overshoot layer, while the turbulent dissipation of the equilibrium tide occurs over the whole convection zone. 

Other possible observational consequences of the mechanism proposed in this work concern the excitation of gravity waves in the stellar radiative zone \citep{Bretonetal22} and the process leading to the inward diffusion and nuclear burning of light elements \citep{MontalbanSchatzman00}. These aspects merit  a dedicated investigation in future works since we have qua\-li\-ta\-ti\-vely predicted a current lower efficiency for both such processes in WASP-18, given the disruption of the downdrafts in the overshoot layer of this star. 

\begin{acknowledgements}
The authors are grateful to Dr.~S.~Mathis for useful discussions and suggestions.  They acknowledge the careful and critical reading by an anonymous Referee that significantly helped to improve the original version of this work. AFL acknowledges support from INAF through their program entitled "Unveiling the magnetic side of the stars" (P.I.~Dr.~A.~Bonanno) financed by a theory grant approved as part of the INAF initiative to foster Fundamental Astrophysics. 
\end{acknowledgements}


\appendix
\section{Downdraft diffusion under the action of turbulent diffusivity}
\label{appendixA}

In the presence of turbulent diffusivity in the overshoot layer as {we expect} to be produced by the equilibrium tidal flow, the lifetime of the downdrafts is affected and their decay is faster than  their intrinsic timescale $\tau_{\rm p}$ as set by the convective flow. 

In order to estimate the effect of the turbulent diffusivity, we assume that the decay of the velocity field of the downdraft is ruled by a diffusion equation to which we add the contribution due to the intrinsic decay of the convective flow, that is,
\begin{equation}
    \frac{\partial \varv}{\partial t} = \nu_{\rm turb\, h} \nabla_{\rm h}^{2} \varv - \frac{\varv}{\tau_{\rm p}},
\label{diffeq}
\end{equation}
where we consider only the effect of the horizontal turbulent diffusivity as parameterised by $\nu_{\rm turb\, h}$ because the turbulent vertical diffusivity is hampered by the sub-adiabatic stratification in the overshoot layer of WASP-18 (cf. Sect.~\ref{turbulent_diff}). Considering that the downdraft is symmetric around its vertical axis \citep[cf.][]{RieutordZahn95}, the horizontal Laplacian is given by:
\begin{equation}
    \nabla_{\rm h}^{2} = \frac{1}{s} \frac{\partial}{\partial s} \left(s \frac{\partial }{\partial s} \right).
    \label{diff_vel}
\end{equation}
We look for a solution of Eq.~\eqref{diffeq} of the form:
\begin{equation}
\varv = g(t) \ \varv_{\rm down} (s, z, t),
\label{vdiff_sol}
\end{equation}
where $g(t)$ is a function of the time that accounts for the effect of the turbulent diffusivity only and $\varv_{\rm down}$ is the downdraft velocity field in the absence of diffusivity as  given by Eq.~\eqref{downdraft_vfield}. 

Rigorously speaking, a solution of the form presented in Eq. \eqref{vdiff_sol} is not allowed in the case of Eq.~\eqref{diffeq}, except along the axis of symmetry of the downdraft, that is, for $s=0$, as can be immediately verified by substituting the trial solution from Eq. \eqref{vdiff_sol} into the same equation. Therefore, limiting ourselves to a solution along the axis $s=0$, we substitute Eq.~\eqref{vdiff_sol} into Eq.~\eqref{diffeq}. Thus, we obtain: 
\begin{equation}
    \frac{dg}{dt} =  -\left(\frac{2\nu_{\rm turb\, h}}{b^{2}} \right) \, g(t),
\end{equation}
which gives
\begin{equation}
g(t) = \exp \left( -\frac{t}{\tau_{\rm d}}\right),
\label{decay_turbul}
\end{equation}
where $\tau_{\rm d} = b^{2}/(2\nu_{\rm turb\, h})$ is the turbulent diffusion timescale. Given that our model provides only an order-of-magnitude estimate, we adopt the turbulent diffusion time $\tau_{\rm d}$ as given by Eq.~\eqref{time_dect}. When we combine the intrinsic decay of the downdraft with the timescale $\tau_{\rm p}$ (cf. Eq.~\ref{downdraft_vfield}) with the additional turbulent decay as given by the solution \eqref{decay_turbul}, we obtain a time dependence of its flow proportional to $ \exp (-t/ \tau_{\rm d}) \exp (-t/ \tau_{\rm p})$, that gives the expression for the lifetime of the downdraft in  Eq.~\eqref{total_diff_time}.

\section{Instabilities in a continuous magnetic layer}
\label{app_continuous_layer}

{Our approach assuming that the magnetic field in the overshoot layer is organised as a set of discrete slender flux tubes has the advantage of a great simplification in the treatment of the effects of  field curvature,  external stratification, and differential rotation, but it cannot include important physical effects such as the doubly diffusive instability that can destabilise a continuous magnetic layer, even in the presence of a strong sub-adiabatic stratification \citep{SchmittRosner83,HughesBrummell21}. A discussion of this effect  is provided by \citet{Hughes07}, for instance, and we  follow this approach in view of their simplicity and focusing on the basic physics. Neglecting the effect of rotation and assuming that there is no bending of the magnetic field lines, the criterion for the onset of the doubly diffusive instability is:
\begin{equation}
    -\frac{g V_{\rm A}^{2}}{c_{\rm s}^{2}} \frac{d}{dz^{\prime}} \ln B > \frac{\eta}{\kappa} N^{2},
    \label{doubly_diff}
\end{equation}
where $V_{\rm A}$ is the Alfven velocity, $c_{\rm s}$ the sound speed, $g$ the acceleration of gravity, $z^{\prime}$ the vertical coordinate increasing toward the stellar surface, $\eta$ the magnetic diffusivity, $\kappa$ the thermal conductivity, and $N$ the Brunt-V\"ais\"ala frequency. In the radiative zone of a star, $\eta \ll \kappa$; therefore, a small perturbation of the magnetic field can exchange heat and reach the temperature of its surrounding, while its magnetic field diffuses much more slowly. This maintains the unstable field gradient responsible for the growth of the perturbation and keeps the field buoyant during its motion, thus increasing the amplitude of the initial perturbation. Given that the ratio of the microscopic diffusivities $\eta/\kappa \sim 10^{-4}$ in the upper radiative zone of the Sun \citep{BrunZahn06}, even a strongly stable stratification, corresponding to a large value of the Brunt-V\"ais\"ala frequency $N$, cannot prevent the instability in the presence of a modest magnetic field gradient. 

In the overshoot layer, the presence of a mild turbulence implies that the microscopic values of $\eta$ and $\kappa$ should be substituted by their turbulent values in a mean-field description of the system. 
In other words, if the Alfven velocity corresponding to the magnetic field is smaller than the turbulent velocity, the turbulent diffusivities have the values derived from the mixing-length theory and their ratio $\eta/\kappa \sim 1$, so that any  doubly diffusive instability is virtually suppressed  \citep[cf.][]{SchmittRosner83,Hughes07}. 

On the other hand, in the presence of a strong magnetic field, we expect the maximum impact on our model because the turbulent diffusivities are strongly quenched, become anisotropic, and strongly dependent on the value of the field itself. Their values could be predicted by numerical models that, however, are many orders of magnitude away from the hydromagnetic regimes characteristic of stars \citep[cf.][]{Kapylaetal20}. In view of the simplifications already introduced in Eq.~\eqref{doubly_diff}, we adopt the analytic first-order theory proposed by \citet{Kitchatinovetal94} and use it to compute the quenching of $\eta$ and $\kappa$ by a magnetic field when the Alfven velocity $V_{\rm A}$ is larger than the turbulent velocity $u_{\rm t}$. We  chose to limit our analysis to the isotropic components of the corresponding $\eta$ and $\kappa$ tensors in the regime $V_{\rm A}/u_{\rm t} \gg 1$ and conservatively adopt the maximum vertical velocity of the downdrafts as the order of magnitude of the turbulent velocity $u_{\rm t}$.  For the specific parameter values we adopt for WASP-18 (see Sects.~\ref{stellar_model} and~\ref{model_application}), we  find that the mean value of the magnetic field at the threshold for the doubly diffusive instability increases by a factor of $\sim 5$ when we take into account the increase in the sub-adiabaticity produced by the reduction in the filling factor of the downdrafts due to the equilibrium tide. Therefore, we confirm the result found with the slender flux tube model that a reduction in the downdraft filling factor leads to a remarkably higher threshold for the instability of the field in the overshoot layer. 

%The ratio of the turbulent diffusivities $\eta/\kappa \sim 0.6 $ for the strongest field value, therefore, the deviation from the fully turbulent case ($\eta/\kappa \sim 1$) is in general not large. 
%This implies that also  overstable oscillations should have a limited impact on our model because their growth rates are maximum when the diffusivities have their microscopic values. 
Another destabilising mechanism that can act on a magnetic layer is represented by overstable oscillations. They are triggered by an increase in the field intensity with height when $\eta < \kappa$ and the density stratification across the magnetic layer can be neglected \citep{Hughes07}. Therefore, they are likely not to be relevant for our model because they require a field gradient that is opposite to that established by the turbulent pumping by the downdrafts that pushes the field downwards thus tending to amplify it towards greater depths. 

A final point to be considered is the influence of the tidal shear on the stability of our continuous magnetic field in the overshoot layer or, from a more general point of view, of a toroidal field in the bulk of the convection zone. Unfortunately,  the magnetic field instabilities produced by a shear flow parallel to the field and having a velocity gradient in the orthogonal direction are functions of  the specific dependence of the velocity on the depth \citep[cf.][]{Hughes07}, so that no general criterion can be found. Nevertheless, the shear associated with the tidal flow in WASP-18, that is, $\partial \dot{\xi}_{\phi}/\partial r$, is about three orders of magnitude smaller than the shear associated with the stellar differential rotation, assuming a radial gradient of the stellar angular velocity in its overshoot layer similar to that of the solar tachocline. A similar conclusion is reached in the case of the convective flows in the bulk of the convection zone. Therefore, we conclude that the effect of the tidal shear on the field stability is likely to be negligible.}

\end{document}